\documentclass[a4paper,12pt]{article}

\pdfoutput=1

\usepackage[tbtags]{amsmath}
\usepackage{amssymb}
\usepackage{ifthen}
\usepackage{slashed}
\usepackage{amsfonts}
\usepackage{mathrsfs}
\usepackage{bm}
\usepackage{graphicx,subfigure,booktabs}
\usepackage[numbers,sort&compress]{natbib}
\usepackage{verbatim}
\usepackage{appendix}

\usepackage[colorlinks,
            linkcolor=black,
            filecolor=black,
            anchorcolor=black,
            urlcolor=black,
            citecolor=blue
            ]{hyperref}

\usepackage{color}
\usepackage{ulem}
\usepackage{setspace}
\numberwithin{equation}{section}

\newlength{\dinwidth}
\newlength{\dinmargin}
\makeatletter
\newcommand{\thickhline}{%
    \noalign {\ifnum 0=`}\fi \hrule height 1pt
    \futurelet \reserved@a \@xhline
}
\makeatother

\allowdisplaybreaks

\begin{document}

\title{}

\title{\bf QCD analysis of the $P$-wave charmonium electromagnetic Dalitz decays \boldmath{$h_{c}\rightarrow\eta^{(\prime)}\ell^{+}\ell^{-}$}}

\author{Chao-Jie Fan\; and
Jun-Kang He\footnote{hejk@hbnu.edu.cn}\\[15pt]
{\small College of Physics and Electronic Science, Hubei Normal University, Huangshi 435002, China}}
\date{}


\maketitle
\vspace{0.2cm}

\begin{abstract}
{\noindent} The $P$-wave charmonium electromagnetic Dalitz decays $h_{c}\rightarrow\eta^{(\prime)}\ell^{+}\ell^{-}$ $(\ell=e, \mu)$ with large recoil momentum are investigated in the framework of perturbative QCD, and the contributions from the small recoil momentum region are described by the overlap of soft wave functions. The transition form factors $f_{h_{c}\eta^{(\prime)}}(q^{2})$ and the normalized transition form factors $F_{h_{c} \eta^{(\prime)}}(q^{2})$ in full kinematic region are derived for the first time. It is noticed that there are no IR divergences at one-loop level, and the transition form factors with the relativistic corrections from the internal momentum of $h_{c}$ are insensitive to both the shapes of $\eta^{(\prime)}$ distribution amplitudes and the invariant mass of the lepton pair in the large recoil momentum region. Intriguingly, unlike the situation in the $S$-wave charmonium decays $J/\psi\rightarrow\eta^{(\prime)}\ell^{+}\ell^{-}$, we find the contributions from the small recoil momentum region are comparable with those from the large recoil momentum region in the $P$-wave charmonium decays $h_{c}\rightarrow\eta^{(\prime)}\ell^{+}\ell^{-}$. By employing the obtained $F_{h_{c} \eta^{(\prime)}}(q^{2})$, we give the predictions of the branching ratios $\mathcal{B}(h_{c}\rightarrow\eta^{(\prime)}\ell^{+}\ell^{-})$, which may come within the range of measurement of present or near-future experiments.
\end{abstract}


\newpage

\section{Introduction}
\label{sec:intro}

The electromagnetic (EM) Dalitz decays of charmonia have received a great deal of attention in the last decade both experimentally~\cite{BESIII:2014dax,BESIII:2018iig,BESIII:2018aao,BESIII:2018qzg,BESIII:2019yeu,BESIII:2019ldo,BESIII:2021xoh,BESIII:2022jde,BESIII:2023jyg} and theoretically~\cite{Fu:2011yy,Chen:2014yta,Kubis:2014gka,Gu:2019qwo,Zhang:2019xia,He:2020jvj,Yan:2023nqz}, since they provide an ideal platform to probe the intrinsic structure of the charmonia and to study the fundamental mechanisms of the interactions between photons and hadrons~\cite{Landsberg:1985kjr,Landsberg:1985gaz}. One of the most interesting topics related to these EM Dalitz decays is the decay of charmonia to the mesons $\eta^{(\prime)}$, since it is directly related to the issue of $\eta-\eta^{\prime}$ mixing, which could offer new opportunities to study the $U(1)_{A}$ anomaly~\cite{Adler:1969gk,Bell:1969ts,Weinberg:1975ui,Witten:1978bc,Witten:1979vv,Veneziano:1979ec,Feldmann:1998vh,Feldmann:1999uf,Escribano:2020jdy} and the $SU(3)_{F}$ breaking~\cite{Kazi:1975tu,Fritzsch:1976qc,Feldmann:1998vh,Feldmann:1999uf,Escribano:2020jdy}. Under the classic assumption of pointlike particles~\cite{Kroll:1955zu,Landsberg:1985gaz}, the EM Dalitz decays $J/\psi (\psi^{\prime})\rightarrow\eta^{(\prime)}\ell^{+}\ell^{-}$ can be described by quantum electrodynamics (QED). And the transition form factors (TFFs) $f_{\psi\eta^{(\prime)}}(q^{2})$, which reflect the deviation from the QED prediction~\cite{Kroll:1955zu,Landsberg:1985gaz}, can provide the dynamical information of the EM structure arising at the $J/\psi(\psi(2S)) \to \eta^{(\prime)}$ transition vertex. Consequently, the TFFs $f_{\psi\eta^{(\prime)}}(q^{2})$ may help to distinguish the transition mechanisms based on different dynamical picture, such as the simple pole approximation~\cite{Fu:2011yy,Gu:2019qwo,Zhang:2019xia}, the effective Lagrangian approach~\cite{Chen:2014yta,Yan:2023nqz}, dispersion theory~\cite{Kubis:2014gka}, and the quantum chromodynamics (QCD) analysis~\cite{He:2020jvj}.

The $P$-wave charmonium $h_{c}$ cannot be directly produced in $e^{+}e^{-}$ collisions because of the quantum numbers $J^{PC} = 1^{+-}$, but it can be produced through $\psi(2S) \to \pi^{0}h_{c}$~\cite{BESIII:2010gid,BESIII:2022tfo}. In recent years, many more decay modes of $h_{c}$ have been searched for at BESIII~\cite{Ablikim:2016uoc,BESIII:2018icg,BESIII:2018qrv,BESIII:2020fsz,BESIII:2021ktv,BESIII:2022olx,BESIII:2022tfo}. Due to the negative $C$ parity, the $h_{c}$ most likely decays into a photon and a pseudoscalar meson $\eta_{c}$ or $\eta$ ($\eta^{\prime}$), in which the radiative decays $h_{c}\rightarrow \gamma\eta^{(\prime)}$ have first been observed by the BESIII Collaboration using about $0.4$ billion $\psi(2S)$ events~\cite{Ablikim:2016uoc}. So far, there are around $3$ billion $\psi(2S)$ events collected with the BESIII detector~\cite{BESIII:2020nme,Ablikim:2023pjg,BESIII:2023lcc}, and it represents about an order-of-magnitude increase in statistics. This provides a good opportunity to study the EM Dalitz decays $h_{c}\rightarrow\eta^{(\prime)}\ell^{+}\ell^{-}$, and their branching ratios can be reached by present or near future experiments, especially for the $\eta^{\prime}$ channels (reaching $10^{-5}$). These EM Dalitz decays could not only offer useful information to constrain theoretical models (as mentioned above) in the charmonium region, but also shed light on the transition mechanism of $h_{c} \to \eta^{(\prime)}$ and the $\eta-\eta^{\prime}$ mixing effects~\cite{Gilman:1987ax,Ball:1995zv,Feldmann:1998vh} in different kinematic regions. Besides, it is more interesting for the $P$-wave charmonia decays. Generally, inclusive $P$-wave charmonia decays suffer from IR divergences in the color-singlet state contributions with the zero-binding approximation~\cite{Barbieri:1976fp,Barbieri:1980yp,Barbieri:1981xz}; while the similar IR divergences do not appear in the exclusive $P$-wave charmonia decays~\cite{Kroll:1997vt,Wong:1998rv,Wong:1999dj,Wong:2000rj}. This may imply that the effects beyond those contained in the derivative of the nonrelativistic wave function at the origin play a key role. Recently, it has been pointed out that the relativistic corrections from the internal momentum of $h_{c}$ are extremely important in the decays $h_{c}\rightarrow \gamma\eta^{(\prime)}$~\cite{He:2020kin}. This indicates that the relativistic corrections may also be important in the EM Dalitz decays $h_{c}\rightarrow\eta^{(\prime)}\ell^{+}\ell^{-}$ due to the same EM structure arising at the $h_{c} \to \eta^{(\prime)}$ transition vertex.

In this paper, one of the major concerns is to clarify the dynamical picture of the EM Dalitz decays $h_{c}\rightarrow\eta^{(\prime)}\ell^{+}\ell^{-}$ in different kinematic regions. Phenomenologically, there exist three types of contributions in the decay processes $h_{c}\rightarrow\eta^{(\prime)}\ell^{+}\ell^{-}$: (i) In the large recoil momentum region, i.e., the square of the invariant mass of the lepton pair $q^{2}=m_{\ell^{+}\ell^{-}}^{2}\simeq 0$, the transition mechanism of $h_{c} \to \eta^{(\prime)}$ could be described by the perturbative QCD approach, which has been reliably employed to treat the corresponding radiative decays $h_{c} \rightarrow\gamma\eta^{(\prime)}$~\cite{Fan:2019sap,He:2020kin}. And we call this transition mechanism the hard mechanism. (ii) In the small recoil momentum region, i.e., the square of the invariant mass of the lepton pair $q^{2}\simeq q^{2}_{\text{max}}=(M_{h_{c}}-m_{\eta^{(\prime)}})^{2}$, the transition mechanism of $h_{c} \to \eta^{(\prime)}$ is governed by the overlapping integration of the soft wave functions, and this transition mechanism is the so-called wave function overlap~\cite{Amsler:1995td,Radyushkin:1998rt,Radyushkin:1998vb,Feldmann:1999sm,Close:2000yk,Huang:2000kd,Chang:2001pm,Li:2007ky,Zhao:2010mm}(i.e., the soft mechanism). The TFFs $f_{h_{c}\eta^{(\prime)}}(q^{2})$ account for the size effects from the spatial wave functions of the initial- and final-state hadrons. (iii) In resonance regions, such as $q^{2}\simeq m_{\rho}^{2}, \, m_{\omega}^{2}, \, m_{\phi}^{2}$, the transition mechanism of $h_{c} \to \eta^{(\prime)}$ can be universally described by the vector meson dominance (VMD) model~\cite{Landsberg:1986fd}, in which the resonance interaction between photons and hadrons is predominant. However, on the one hand the contributions from VMD are negligibly small due to the narrow widths of resonances (see Ref.~\cite{He:2020jvj} for more details), and on the other hand there are still some open questions for the VMD model, such as the sign ambiguity in the amplitude from the intermediate vector mesons and the off-mass-shell effects of the coupling constants~\cite{Intemann:1983yj}. To make the dynamical picture of the EM Dalitz decays $h_{c}\rightarrow\eta^{(\prime)}\ell^{+}\ell^{-}$ clear, we will mainly present the detailed discussions about the hard mechanism and the soft mechanism in the later parts of this paper. In the large recoil momentum region, by employing the Bethe-Salpeter (B-S) framework~\cite{Salpeter:1951sz,Salpeter:1952ib,Mitra:1990av,Bhatnagar:2009jg,Bhatnagar:2013bha,He:2020kin}, we work out the B-S wave function of $h_{c}$, in which the internal momentum is retained. Considering a large momentum transfer, one can adopt the light-cone distribution amplitudes (DAs) to describe the internal dynamics of the final light mesons $\eta^{(\prime)}$, and the involved quark-antiquark and gluonic contents of $\eta^{(\prime)}$ are taken into account in our calculations. By an analytic calculation of the involved one-loop integrals, we find that the TFFs $f_{h_{c}\eta^{(\prime)}}(q^{2})$ are UV and IR safe, and they barely depend on the shapes of the light meson DAs. Furthermore, the gluonic contributions and the quark-antiquark contributions are comparable in the TFFs $f_{h_{c}\eta^{(\prime)}}(q^{2})$. It is compatible with the situation in the corresponding radiative decays $h_{c} \rightarrow\gamma\eta^{(\prime)}$~\cite{Fan:2019sap,He:2020kin}. In the small recoil momentum region, the TFFs are calculated phenomenologically by the wave function overlap. Through a detailed calculation, we obtain the TFFs $f_{h_{c}\eta^{(\prime)}}(q^{2})$ in the whole kinematic region for the first time. It is worthwhile to point out that the contributions from the soft mechanism and those from the hard mechanism are comparable with each other in the branching ratios $\mathcal{B}(h_{c}\rightarrow\eta^{(\prime)}\ell^{+}\ell^{-})$, unlike the situation in $S$-wave charmonium EM Dalitz decays $J/\psi\rightarrow\eta^{(\prime)}\ell^{+}\ell^{-}$~\cite{He:2020jvj} where the soft contributions are suppressed because of the special form of the spin structure of their amplitudes. In order to remove the main uncertainties arising from the bound-state wave functions, we use the normalized TFFs $F_{h_{c} \eta^{(\prime)}}(q^{2})\equiv f_{h_{c}\eta^{(\prime)}}(q^{2})/f_{h_{c}\eta^{(\prime)}}(0)$ to obtain the predictions of the branching ratios $\mathcal{B}(h_{c}\rightarrow\eta^{(\prime)}\ell^{+}\ell^{-})$.

The paper is organized as follows: The theoretical framework for the EM Dalit decays $h_{c}\rightarrow\eta^{(\prime)}\ell^{+}\ell^{-}$ is presented in detail in Sec.~\ref{sec:framework}. In Sec.~\ref{sec:Results and discussions} we show our numerical results and some phenomenological discussions, and Sec.~\ref{sec:conclusion} is our summary.

\section{THEORETICAL FRAMEWORK}
\label{sec:framework}
\subsection{Hard mechanism}
\label{subsec:Hard mechanism}
\subsubsection{Contributions of the quark-antiquark content of $\eta^{(\prime)}$}
\label{subsubsec:QCDq}

\begin{figure}[th]
  \begin{center}
  \includegraphics[width=0.4\textwidth]{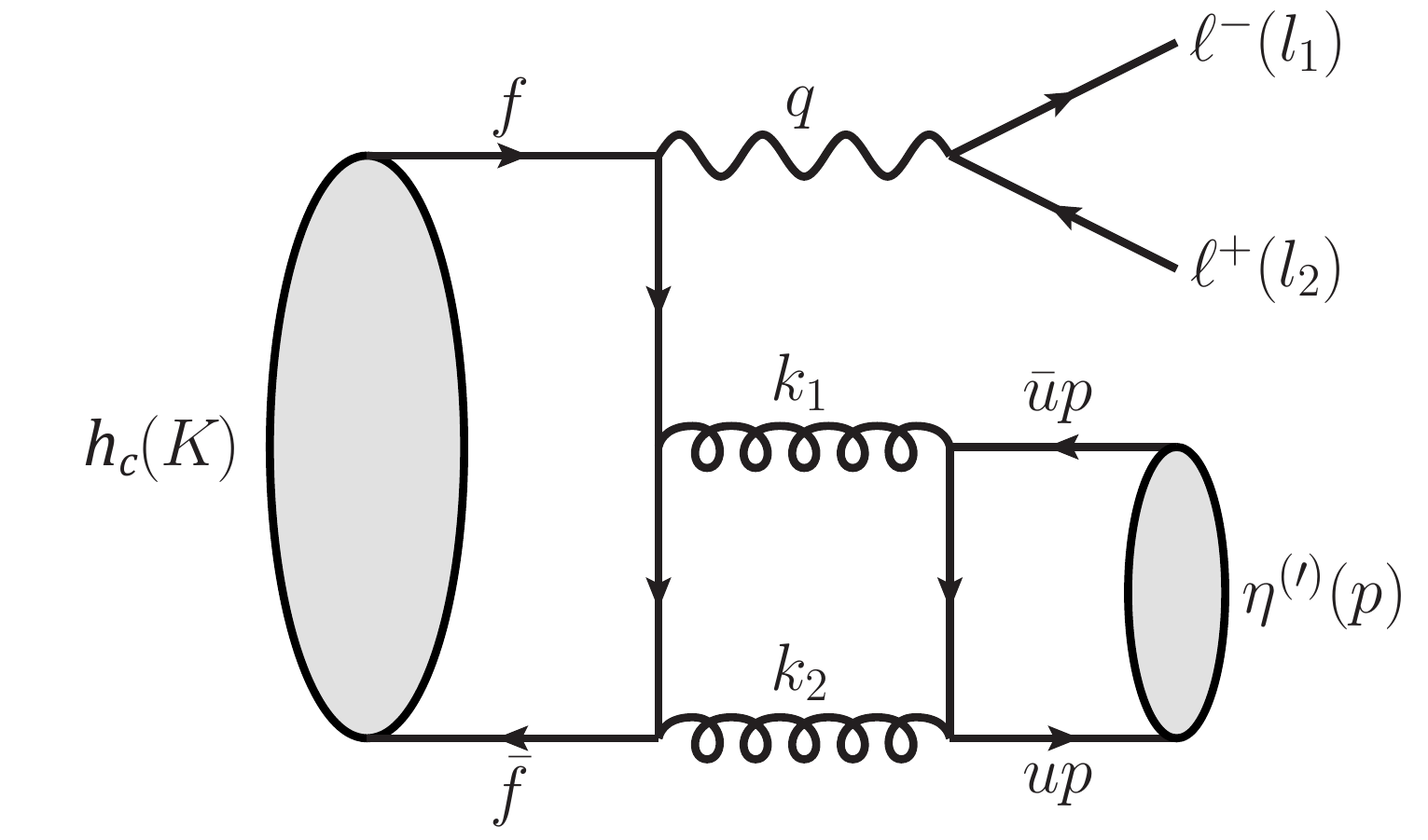}
  \end{center}
  \vskip -0.7cm
  \caption{One typical Feynman diagram for $h_{c}\rightarrow\eta^{(\prime)}\ell^{+}\ell^{-}$ with the quark-antiquark content of $\eta^{(\prime)}$, and the kinematical variables are labeled.}\label{QCDq}
\end{figure}

In the large recoil momentum region of $\eta^{(\prime)}$, the EM Dalitz decays $h_{c}\rightarrow\eta^{(\prime)}\ell^{+}\ell^{-}$ can be described by the perturbative QCD approach. The leading order Feynman diagrams for the quark-antiquark content of $\eta^{(\prime)}$ arise from one-loop QCD processes. One of them is illustrated in Fig.~\ref{QCDq}, and the other five diagrams come from permutations of the photon and the gluon legs. Here and in what follows, the involved kinematical variables are labeled in Fig.~\ref{QCDq}, where $u$ and $\bar{u}$ are the momentum fractions carried by the light quark and the light antiquark, respectively. According to the Feynman diagrams, one can obtain the amplitude of $h_{c}\rightarrow\eta^{(\prime)}\ell^{+}\ell^{-}$,
\begin{eqnarray}
{\mathcal M}=-\frac{e}{q^{2}}{\mathcal A}^{\alpha\beta}\varepsilon_{\alpha}(K)\bar{u}(l_{1})\gamma_{\beta} v(l_{2}),
\end{eqnarray}
where $\mathcal{A}^{\alpha\beta}$ represents the amplitude of $h_{c}\rightarrow\eta^{(\prime)}\gamma^{\ast}$; $K$ and $\varepsilon(K)$ stand for the momentum and polarization vector of $h_{c}$, respectively; $q$ stands for the momentum of the virtual photon, and $q^{2}=m_{\ell^{+}\ell^{-}}^{2}$ is the square of the invariant mass of the lepton pair; $l_{1}$ and $l_{2}$ stand for the momenta of the leptons $\ell^{-}$ and $\ell^{+}$, respectively.

It is convenient to convert the amplitude of $h_{c}\rightarrow \eta^{(\prime)}\gamma^{\ast}$ into two parts: the effective coupling of the process $h_{c}\rightarrow g^{\ast}g^{\ast} \gamma^{\ast}$ and that of the process $g^{\ast}g^{\ast}\rightarrow\eta^{(\prime)}$. By multiplying the two parts, inserting the gluon propagators and performing the loop integrations, one can obtain the final amplitude of $h_{c}\rightarrow \eta^{(\prime)}\gamma^{\ast}$.

In the rest frame of the $P$-wave charmonium $h_{c}$, the amplitude of $h_{c}\rightarrow g^{\ast}g^{\ast} \gamma^{\ast}$ can be written as~\cite{Guberina:1980dc,Guberina:1980xb,Korner:1982vg}
\begin{eqnarray}\label{ggr}
{\mathcal A}^{\alpha\beta\mu\nu}_{1}\varepsilon_{\alpha}(K)\epsilon^{*}_{\beta}(q)\epsilon^{*}_{\mu}(k_{1})\epsilon^{*}_{\nu}(k_{2})
&=&\sqrt{3}\int\frac{\mathrm{d}^{4}k}{(2\pi)^{4}}\textrm{Tr}\left[\Psi(K,k){\cal O}(f,\bar{f})\right],
\end{eqnarray}
where $\Psi(K,k)$ represents the B-S wave function of $h_{c}$; $\mathcal{O}(f,\bar{f})$ represents the hard-scattering amplitude; $\sqrt{3}$ represents the color factor; $\epsilon(q)$ represents the polarization vector of the virtual photon; $k_{1}$, $k_{2}$ and $\epsilon(k_{1})$, $\epsilon(k_{2})$ represent the two gluons' momenta and polarization vectors; $f$ and $\bar{f}$ represent the momenta of the quark $c$ and antiquark $\bar{c}$, and they read
\begin{eqnarray}
f^{\mu}=\frac{K^{\mu}}{2}+k^{\mu}=\left(\frac{M}{2}+k^{0},\mathbf{k}\right),\quad\quad   \bar{f}^{\mu}=\frac{K^{\mu}}{2}-k^{\mu}=\left(\frac{M}{2}-k^{0},-\mathbf{k}\right)
\end{eqnarray}
with $k$ the relative momentum between the quark $c$ and antiquark $\bar{c}$, i.e., the internal momentum of the $P$-wave charmonium $h_{c}$. Here $M$ is the mass of $h_{c}$. For convenience in subsequent calculations, we divide the internal momentum of $h_{c}$ into two parts: the transverse component $\hat{k}$ with $\hat{k}\cdot K=0$ and the longitudinal component $k_{\parallel}$ with $k_{\parallel}\cdot \hat{k}=0$, i.e.,
\begin{eqnarray}\label{qdecom}
k^{\mu}= k_{\parallel}^{\mu}+\hat{k}^{\mu},\quad\quad  k_{\parallel}^{\mu} = \frac{k_{K}}{M} K^{\mu},
\end{eqnarray}
where both $k_{K}=\frac{k\cdot K}{M}$ and $\hat{k}^{2}=k^{2}-k_{K}^{2}$ are Lorentz invariant variables. Considering the rest frame of $h_{c}$, one can easily know that $\hat{k}$ involves $3$ degrees of freedom (namely, the component $\mathbf{k}$) orthogonal to the total momentum $K$, and $k_{K}$ contains the remaining $1$ degree of freedom (namely, the component $k^{0}$). Now the volume element of the internal momentum $k$ can be expressed in the form $\mathrm{d}^{4}k=\mathrm{d}^{3}\hat{k}\mathrm{d}k_{K}$. Furthermore, with a more relevant treatment $k^{0}\ll M$, we obtain the momenta
\begin{eqnarray}
f^{\mu}\approx\left(\frac{M}{2},\mathbf{k}\right)=\frac{K^{\mu}}{2}+\hat{k}^{\mu},\quad\quad  \bar{f}^{\mu}\approx\left(\frac{M}{2},-\mathbf{k}\right)=\frac{K^{\mu}}{2}-\hat{k}^{\mu},
\end{eqnarray}
and the hard-scattering amplitude
\begin{eqnarray}\label{hsao}
{\cal O}(f,\bar{f})&\approx&{\cal O}(\hat{k}),
\end{eqnarray}
and this treatment maintains the gauge invariance of the hard-scattering amplitude~\cite{Chao:1995cz}.

By employing the B-S equation~\cite{Salpeter:1951sz,Salpeter:1952ib} of the $P$-wave charmonium $h_{c}$, one can reduce the B-S equation to the Salpeter equation under the covariant instantaneous ansatz (CIA)~\cite{Mitra:1990av,Bhatnagar:2009jg,Bhatnagar:2013bha}. The Salpeter wave function is defined as
\begin{equation}\label{bssalfun}
\psi(\hat{k})=\frac{i}{2\pi}\int\mathrm{d}k_{K}\Psi(K,k).
\end{equation}
Subsequently, we obtain an analytic Salpeter wave function of $h_{c}$ by solving the Salpeter equation (more details can be found in our recent investigation~\cite{He:2020kin}),
\begin{eqnarray}\label{bsfun}
\psi(\hat{k})=\hat{k}\cdot\varepsilon(K) \left[1+\frac{\slashed{K}}{M}+\frac{\hat{\slashed{k}}\slashed{K}}
{\hat{m}_{c}M}\right]\gamma^{5}f(\hat{k}^{2}),
\end{eqnarray}
where $\hat{m}_{c}$ is the effective mass of $c$ quark, and the front factor $\hat{k}\cdot\varepsilon(K)$ indicates that the wave function is the nature of $P$-wave, and the scalar function $f(\hat{k}^{2})$  reads
\begin{eqnarray}\label{fhatq}
f(\hat{k}^{2})&=&N_{A}\left(\frac{2}{3}\right)^{\frac{1}{2}}\frac{1}{\pi^{\frac{3}{4}}\beta^{\frac{5}{2}}_{{A}}}
|\mathbf{\hat{k}}|e^{-\frac{\mathbf{\hat{k}}^{2}}{2\beta^{2}_{A}}}
\end{eqnarray}
with $N_{A}$ the normalization constant and $\beta_{A}$ the harmonic oscillator parameter. And the normalization equation of  $f(\hat{q}^{2})$ reads
\begin{eqnarray}
\int\frac{\mathrm{d}^{3}\hat{k}}{(2\pi)^{3}}\frac{4\omega \mathbf{\hat{k}}^{2}}{3\hat{m}_{c} M}f^{2}(\hat{k}^{2})=1.
\end{eqnarray}

Using Eqs.~\eqref{hsao} and~\eqref{bssalfun}, we can rewrite the amplitude of $h_{c}\rightarrow\gamma g^{\ast}g^{\ast}$:
\begin{eqnarray}
{\mathcal A}^{\alpha\beta\mu\nu}_{1}\varepsilon_{\alpha}(K)\epsilon^{*}_{\beta}(q)\epsilon^{*}_{\mu}(k_{1})\epsilon^{*}_{\nu}(k_{2})
=-i\sqrt{3}\int\frac{\mathrm{d}^{3}\hat{k}}{(2\pi)^{3}}\textrm{Tr}\left[\psi
(\hat{k}){\cal O}(\hat{k})\right],
\end{eqnarray}
where the hard-scattering amplitude ${\cal O}(\hat{k})$ reads
\begin{eqnarray}
{\cal O}(\hat{k})=iQ_{c}eg^{2}_{s}
\frac{\delta_{ab}}{6}&&\Bigg{[}\slashed{\epsilon}^{\ast}(k_{2})\frac{\frac{\slashed{k}_{2}-\slashed{q}-\slashed{k}_{1}}{2}
+\hat{\slashed{k}}+m_{c}}{\left(\frac{k_{2}-q-k_{1}}{2}+\hat{k}\right)^{2}-m^{2}_{c}}
\slashed{\epsilon}^{\ast}(q)\frac{\frac{\slashed{k}_{2}
+\slashed{q}-\slashed{k}_{1}}{2}+\hat{\slashed{k}}+m_{c}}
{\left(\frac{k_{2}+q-k_{1}}{2}+\hat{k}\right)^{2}-m^{2}_{c}}\slashed{\epsilon}^{\ast}(k_{1})\nonumber\\
& &+\slashed{\epsilon}^{\ast}(k_{1})\frac{\frac{\slashed{k}_{1}-\slashed{q}-\slashed{k}_{2}}{2}
+\hat{\slashed{k}}+m_{c}}{\left(\frac{k_{1}-q-k_{2}}{2}+\hat{k}\right)^{2}-m^{2}_{c}}
\slashed{\epsilon}^{\ast}(q)\frac{\frac{\slashed{k}_{1}
+\slashed{q}-\slashed{k}_{2}}{2}+\hat{\slashed{k}}+m_{c}}
{\left(\frac{k_{1}+q-k_{2}}{2}+\hat{k}\right)^{2}-m^{2}_{c}}\slashed{\epsilon}^{\ast}(k_{2})\nonumber\\
& &+\slashed{\epsilon}^{\ast}(k_{2})\frac{\frac{\slashed{k}_{2}-\slashed{k}_{1}-\slashed{q}}{2}
+\hat{\slashed{k}}+m_{c}}{\left(\frac{k_{2}-k_{1}-q}{2}+\hat{k}\right)^{2}-m^{2}_{c}}
\slashed{\epsilon}^{\ast}(k_{1})\frac{\frac{\slashed{k}_{2}
+\slashed{k}_{1}-\slashed{q}}{2}+\hat{\slashed{k}}+m_{c}}
{\left(\frac{k_{2}+k_{1}-q}{2}+\hat{k}\right)^{2}-m^{2}_{c}}\slashed{\epsilon}^{\ast}(q)\nonumber\\
& &+\slashed{\epsilon}^{\ast}(q)\frac{\frac{\slashed{q}-\slashed{k}_{2}-\slashed{k}_{1}}{2}
+\hat{\slashed{k}}+m_{c}}{\left(\frac{q-k_{2}-k_{1}}{2}+\hat{k}\right)^{2}-m^{2}_{c}}
\slashed{\epsilon}^{\ast}(k_{2})\frac{\frac{\slashed{q}
+\slashed{k}_{2}-\slashed{k}_{1}}{2}+\hat{\slashed{k}}+m_{c}}
{\left(\frac{q+k_{2}-k_{1}}{2}+\hat{k}\right)^{2}-m^{2}_{c}}\slashed{\epsilon}^{\ast}(k_{1})\nonumber\\
& &+\slashed{\epsilon}^{\ast}(k_{1})\frac{\frac{\slashed{k}_{1}-\slashed{k}_{2}-\slashed{q}}{2}
+\hat{\slashed{k}}+m_{c}}{\left(\frac{k_{1}-k_{2}-q}{2}+\hat{k}\right)^{2}-m^{2}_{c}}
\slashed{\epsilon}^{\ast}(k_{2})\frac{\frac{\slashed{k}_{1}
+\slashed{k}_{2}-\slashed{q}}{2}+\hat{\slashed{k}}+m_{c}}
{\left(\frac{k_{1}+k_{2}-q}{2}+\hat{k}\right)^{2}-m^{2}_{c}}\slashed{\epsilon}^{\ast}(q)\nonumber\\
& &+\slashed{\epsilon}^{\ast}(q)\frac{\frac{\slashed{q}-\slashed{k}_{1}-\slashed{k}_{2}}{2}
+\hat{\slashed{k}}+m_{c}}{\left(\frac{q-k_{1}-k_{2}}{2}+\hat{k}\right)^{2}-m^{2}_{c}}
\slashed{\epsilon}^{\ast}(k_{1})\frac{\frac{\slashed{q}
+\slashed{k}_{1}-\slashed{k}_{2}}{2}+\hat{\slashed{k}}+m_{c}}
{\left(\frac{q+k_{1}-k_{2}}{2}+\hat{k}\right)^{2}-m^{2}_{c}}\slashed{\epsilon}^{\ast}(k_{2})
\Bigg{]}
\end{eqnarray}
with the $c$ quark mass $m_{c}$.

To proceed, we treat the light mesons $\eta^{(\prime)}$ as a light-cone object in the large recoil momentum region because of a large momentum transfer. Using the light-cone expansion, one can obtain the amplitude of $g^{\ast}g^{\ast}\rightarrow\eta^{(\prime)}$~\cite{Muta:1999tc,Yang:2000ce,Ali:2000ci,He:2019mpy,Fan:2019sap}:
\begin{eqnarray}
\mathcal{A}^{\mu\nu}_{2}=-i (4\pi \alpha_{s})\delta_{ab}\epsilon^{\mu\nu\rho\sigma}k_{1\rho}k_{2\sigma}\sum_{q=u,d,s}\frac{f_{\eta^{(\prime)}}^{q}}{6}
\int^{1}_{0}du\phi^{q}(u)\left(\frac{1}{\bar{u}k_{1}^{2}+uk_{2}^{2}-u\bar{u}m^{2}}+(u\leftrightarrow\bar{u})\right),
\end{eqnarray}
where $m$ represents the mass of $\eta^{(\prime)}$, $f_{\eta^{(\prime)}}^{q}$ are the decay constants, and $\phi^{q}(u)$ is the light-cone DA. The DA can be expressed as~\cite{Agaev:2014wna,Fan:2019sap}
\begin{eqnarray}
\phi^{q}(u)&=&6u(1-u)\left[1+\sum_{n=2,4\cdots}c^{q}_{n}(\mu)C_{n}^{\frac{3}{2}}(2u-1)\right]
\end{eqnarray}
with $c^{q}_{n}(\mu)$ the Gegenbauer moments, and we take three typical models listed in Table 1 of Ref.~\cite{Fan:2019sap} (see Refs.~\cite{Agaev:2014wna,Fan:2019sap} for more details). In our subsequent calculations, it is found that the TFFs $f_{h_{c}\eta^{(\prime)}}(q^{2})$ are insensitive to the models of the light-cone DA. The decay constants $f_{\eta^{(\prime)}}^{q}$, in the quark-flavor basis, can be parametrized as~\cite{Akhoury:1987ed,Ball:1995zv,Feldmann:1998vh,Feldmann:1998sh,Feldmann:1999uf}
\begin{eqnarray}
f_{\eta}^{u(d)}&=&\frac{f_{q}}{\sqrt{2}}\cos\phi,\quad\quad   f_{\eta}^{s}=-f_{s}\sin\phi, \nonumber\\
f_{\eta^{\prime}}^{u(d)}&=&\frac{f_{q}}{\sqrt{2}}\sin\phi,\quad\quad   f_{\eta^{\prime}}^{s}=f_{s}\cos\phi,
\end{eqnarray}
where the phenomenological parameters ($\phi$, $f_{q}$ and $f_{s}$) could be determined by different methods~\cite{Feldmann:1998vh,Gregory:2011sg,Michael:2013gka,Escribano:2013kba,Chen:2014yta,Escribano:2015nra,Escribano:2015yup,Urbach:2017rvx,He:2019mpy,Fan:2019sap,He:2020kin}.

By contracting the above two amplitudes, inserting the gluon propagators and integrating over the loop momentum, we obtain the decay amplitude of $h_{c}\rightarrow \eta^{(\prime)}\gamma^{\ast}$
\begin{eqnarray}\label{aalph}
\mathcal{A}^{\alpha\beta}=\frac{1}{2}\int\frac{\mathrm{d}^{4}k_{1}}{(2\pi)^{4}}{\mathcal A}^{\alpha\beta\mu\nu}_{1}\mathcal{A}_{2\mu\nu}\frac{i}{k^{2}_{1}
+i\epsilon}\frac{i}{k^{2}_{2}+i\epsilon}.
\end{eqnarray}
Considering parity conservation, Lorentz invariance, gauge invariance, and current conservation, one knows
\begin{eqnarray}\label{Amunu}
\mathcal{A}^{\alpha\beta}\propto \left(-g^{\alpha\beta}+\frac{q^{\alpha}K^{\beta}}{q\cdot K}\right).
\end{eqnarray}
Then the $h_{c}\rightarrow \eta^{(\prime)}\gamma^{\ast}$ TFFs can be defined by
\begin{eqnarray}
\mathcal{A}^{\alpha\beta}=-e f^{Q}_{h_{c} \eta^{(\prime)}}(q^{2}) \left(-g^{\alpha\beta}+\frac{q^{\alpha}K^{\beta}}{q\cdot K}\right).
\end{eqnarray}
With the help of the projection operator
\begin{eqnarray}
\mathcal{P}^{\alpha\beta}=\left( 2+\frac{M^{2}q^{2}}{(q\cdot K)^{2}} \right)^{-1}\left(-g^{\alpha\beta}+\frac{q^{\alpha}K^{\beta}}{q\cdot K}\right),
\end{eqnarray}
the TFFs can be rewritten as
\begin{eqnarray}\label{fpsiet}
f^{Q}_{h_{c} \eta^{(\prime)}}(q^{2})=-e^{-1}\mathcal{P}_{\alpha\beta}\mathcal{A}^{\alpha\beta}.
\end{eqnarray}
Here we show the expression of the TFFs more clearly
\begin{eqnarray}\label{analyfpsiet}
f^{Q}_{h_{c} \eta^{(\prime)}}(q^{2})&=&-\frac{i8\pi^2 \alpha_s^2}{9 \sqrt{3}}\sum_{q}f_{\eta^{(\prime)}}^{q}\int\frac{\mathrm{d}^{3}\hat{k}}{(2\pi)^{3}}\int\mathrm{d}u \phi^{q}(u)\int\frac{\mathrm{d}^{4}k_{1}}{(2\pi)^{4}}\bigg{(}\Big{(}\frac{N_{1}}{D_{1}D_{2}D_{3}D_{4}}\nonumber\\
&&+\frac{N_{2}}{D_{1}D_{2}D_{3}D_{4p}}+\frac{N_{3}}{D_{1}D_{2}D_{3}D_{5}}+\frac{N_{4}}{D_{1}D_{2}D_{3}D_{5p}}+\frac{N_{5}}{D_{1}D_{2}D_{3}D_{4p}D_{5}}\nonumber\\
&&+\frac{N_{6}}{D_{1}D_{2}D_{3}D_{4}D_{5p}}\Big{)}+(u \rightarrow \bar{u})\bigg{)}
\end{eqnarray}
with $u$ and $\bar{u}=(1-u)$ the momentum fractions arising from the light mesons $\eta^{(\prime)}$. The expressions of the denominators read
\begin{eqnarray}
D_{1}&=&k_{1}^{2}+i \epsilon ,\nonumber\\
D_{2}&=&(k_{1}-p)^{2}+i \epsilon ,\nonumber\\
D_{3}&=&(k_{1}-u p)^{2}+i \epsilon ,\nonumber\\
D_{4}&=&(k_{1}-\hat{k}+\frac{q-p}{2})^{2}-m_{c}^{2}+i \epsilon ,\nonumber\\
D_{4p}&=&(k_{1}+\hat{k}+\frac{q-p}{2})^{2}-m_{c}^{2}+i \epsilon ,\nonumber\\
D_{5}&=&(k_{1}+\hat{k}-\frac{q+p}{2})^{2}-m_{c}^{2}+i \epsilon ,\nonumber\\
D_{5p}&=&(k_{1}-\hat{k}-\frac{q+p}{2})^{2}-m_{c}^{2}+i \epsilon ,
\end{eqnarray}
and the expressions of the numerators $N_{1}\sim N_{6}$ are presented in the Appendix.

With the help of the algebraic identity~($u\neq 0,1$)
\begin{equation}
\frac{D_{1}}{u m^{2}}+\frac{D_{2}}{(1-u) m^{2}}-\frac{D_{3}}{u (1-u) m^{2}}=1,
\end{equation}
the TFFs $f^{Q}_{h_{c} \eta^{(\prime)}}(q^{2})$ can be decomposed into a sum of four-point and three-point one-loop integrals. When $u = 0,1$, the denominators of the propagators have the relation $D_{3}=D_{1},D_{2}$, and the TFFs can also be decomposed into four-point or three-point integrals. Then one can analytically evaluate these one-loop integrals with the technique proposed in Refs.~\cite{tHooft:1978jhc,Denner:1991qq,Denner:1991kt} or the computer program \textsc{PACKAGE-X}~\cite{Patel:2015tea,Patel:2016fam}. It is found that the TFFs are UV and IR safe. Similarly to the situation in the radiative decays $h_{c} \rightarrow\gamma\eta^{(\prime)}$~\cite{Fan:2019sap,He:2020kin}, the TFFs are insensitive to the DAs of $\eta^{(\prime)}$. Our numerical results show that the change of the modulus of the TFFs does not exceed $1\%$ with the different models of the DAs. Therefore, the theoretical uncertainties from the DAs are ignorable in our calculations of the TFFs, and we choose model I of the meson DA in Table $1$ of Ref.~\cite{Fan:2019sap}.

\subsubsection{Contributions of the gluonic content of $\eta^{(\prime)}$}
\label{subsubsec:QCDg}

\begin{figure}[th]
  \begin{center}
  \includegraphics[width=0.4\textwidth]{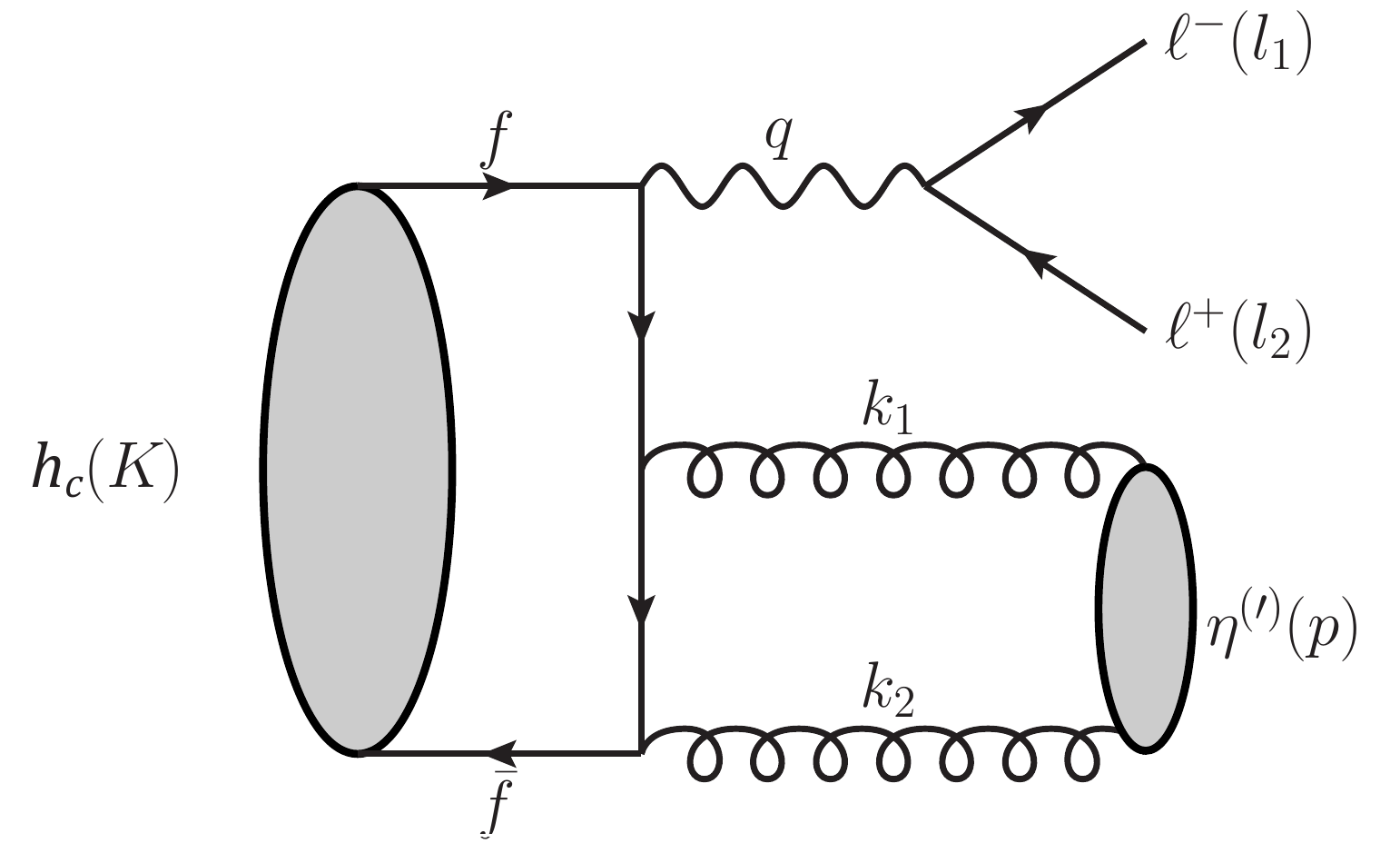}
  \end{center}
  \vskip -0.7cm
  \caption{One typical Feynman diagram for $h_{c}\rightarrow\eta^{(\prime)}\ell^{+}\ell^{-}$ with the gluonic content of $\eta^{(\prime)}$. The kinematical variables are labeled.}\label{QCDg}
\end{figure}

Generally speaking, the contributions of the gluonic content of $\eta^{(\prime)}$ are expected to be small because the gluonic content cloud be seen as higher-order effects from the point of view of the QCD evolution of the two-gluon DA, which vanishes in the asymptotic limit. For example, the gluonic contributions are strongly suppressed by the factor $m_{\eta^{(\prime)}}^{2}/M_{J/\psi}^{2}$ in the radiative decays $J/\psi \rightarrow\gamma\eta^{(\prime)}$~\cite{He:2019mpy}. However, there is no suppression factor in the $P$-wave charmonium radiative decays $h_{c} \rightarrow\gamma\eta^{(\prime)}$~\cite{Fan:2019sap,He:2020kin}, due to the special form of the spin structure in their amplitudes. So the gluonic contributions may become important in the radiative decays of the $P$-wave charmonium $h_{c}$. In fact, as pointed out in Refs.~\cite{Fan:2019sap,He:2020kin}, the gluonic contributions and the quark-antiquark contributions are comparable with each other in the radiative decays $h_{c} \rightarrow\gamma\eta^{(\prime)}$. Obviously, this situation should be found in the large recoil momentum region of $\eta^{(\prime)}$  of the decays $h_{c}\rightarrow\eta^{(\prime)}\ell^{+}\ell^{-}$, due to the same spin structures in their hadronic matrix elements. The corresponding Feynman diagram is depicted in Fig.~\ref{QCDg}, and the other two diagrams arise from permutations of the photon and the gluon legs.

At the leading twist level, the matrix elements of the mesons $\eta^{(\prime)}$ over two-gluon fields in the light-cone expansion can be written as~\cite{Kroll:2002nt,Ball:2007hb,Agaev:2014wna}
\begin{eqnarray}
\langle\eta^{(\prime)}(p)|A_{\alpha}^{a}(x)A_{\beta}^{b}(y)|0\rangle=\frac{1}{4}\epsilon_{\alpha\beta\mu\nu}
\frac{n^{\mu}p^{\nu}}{p\cdot n}\frac{C_{F}}{\sqrt{3}}\frac{\delta^{ab}}{8}f_{\eta^{(\prime)}}^{1}\int\textrm{d}u e^{i(up\cdot x+\bar{u}p\cdot y)}\frac{\phi^{g}(u)}{u(1-u)},
\end{eqnarray}
where $n=\frac{1}{\sqrt{2}}(1,\,-\frac{\mathbf{p}}{|\mathbf{p}|})$ is a lightlike vector along the opposite direction of the mesons $\eta^{(\prime)}$~\cite{Kroll:2002nt}, $f_{\eta^{(\prime)}}^{1}=\frac{1}{\sqrt{3}}(f_{\eta^{(\prime)}}^{u}+f_{\eta^{(\prime)}}^{d}+f_{\eta^{(\prime)}}^{s})$ are the effective decay constant, and the gluonic twist-$2$ DA is~\cite{Agaev:2014wna,Ball:2007hb,Alte:2015dpo}
\begin{eqnarray}
\phi^{g}(u)=30u^{2}(1-u)^{2}\sum_{n=2,4\cdots}c^{g}_{n}(\mu)C_{n-1}^{\frac{5}{2}}(2u-1).
\end{eqnarray}

After a series of calculations, we obtain the corresponding TFFs $f^{G}_{h_{c} \eta^{(\prime)}}(q^{2})$
\begin{eqnarray}\label{fg}
f^{G}_{h_{c} \eta^{(\prime)}}(q^{2}) &=& - \frac{16\pi\alpha_{s}}{27} f_{\eta^{(\prime)}}^{1}\int\mathrm{d}u \frac{\phi^{g}(u)}{u(1-u)}\int\frac{\mathrm{d}^{3}\hat{k}}{(2\pi)^{3}} \left(\frac{N_{7}}{C_{1}C_{2}}+\frac{N_{8}}{C_{2}C_{3}}+\frac{N_{9}}{C_{3}C_{4}}\right),
\end{eqnarray}
where the expressions of the denominators read
\begin{eqnarray}
C_{1}&=&(\hat{k}+q-\frac{K}{2})^{2}-m_{c}^{2}+i \epsilon ,\nonumber\\
C_{2}&=&(\hat{k}-\bar{u}p+\frac{K}{2})^{2}-m_{c}^{2}+i \epsilon ,\nonumber\\
C_{3}&=&(\hat{k}+up-\frac{K}{2})^{2}-m_{c}^{2}+i \epsilon ,\nonumber\\
C_{4}&=&(\hat{k}-q+\frac{K}{2})^{2}-m_{c}^{2}+i \epsilon ,
\end{eqnarray}
and the expressions of the numerators $N_{7}\sim N_{9}$ are presented in the Appendix.

\begin{figure}[!!htb]
\centering
\includegraphics[width=0.43\textwidth]{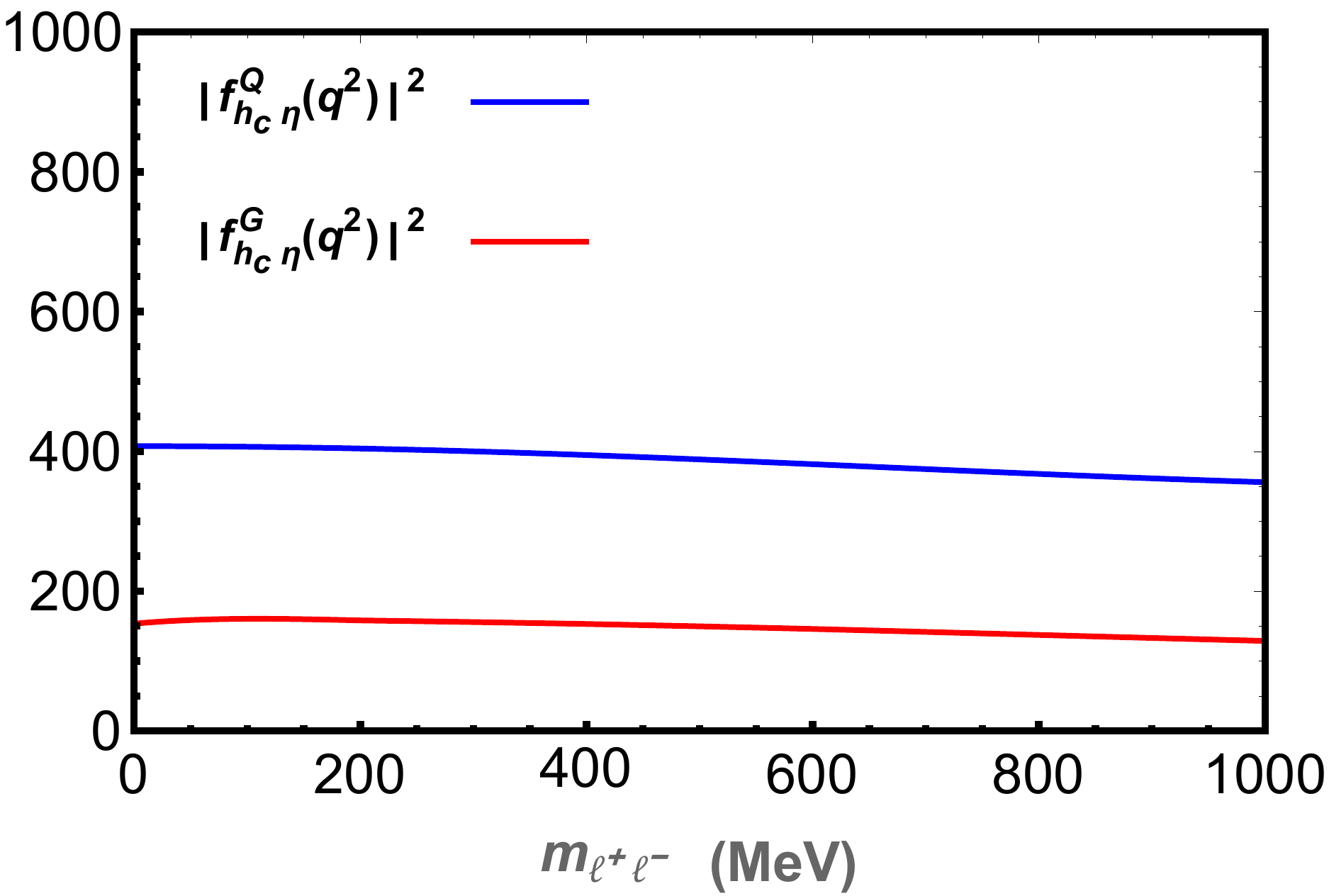}\hspace{0.5cm}
\includegraphics[width=0.43\textwidth]{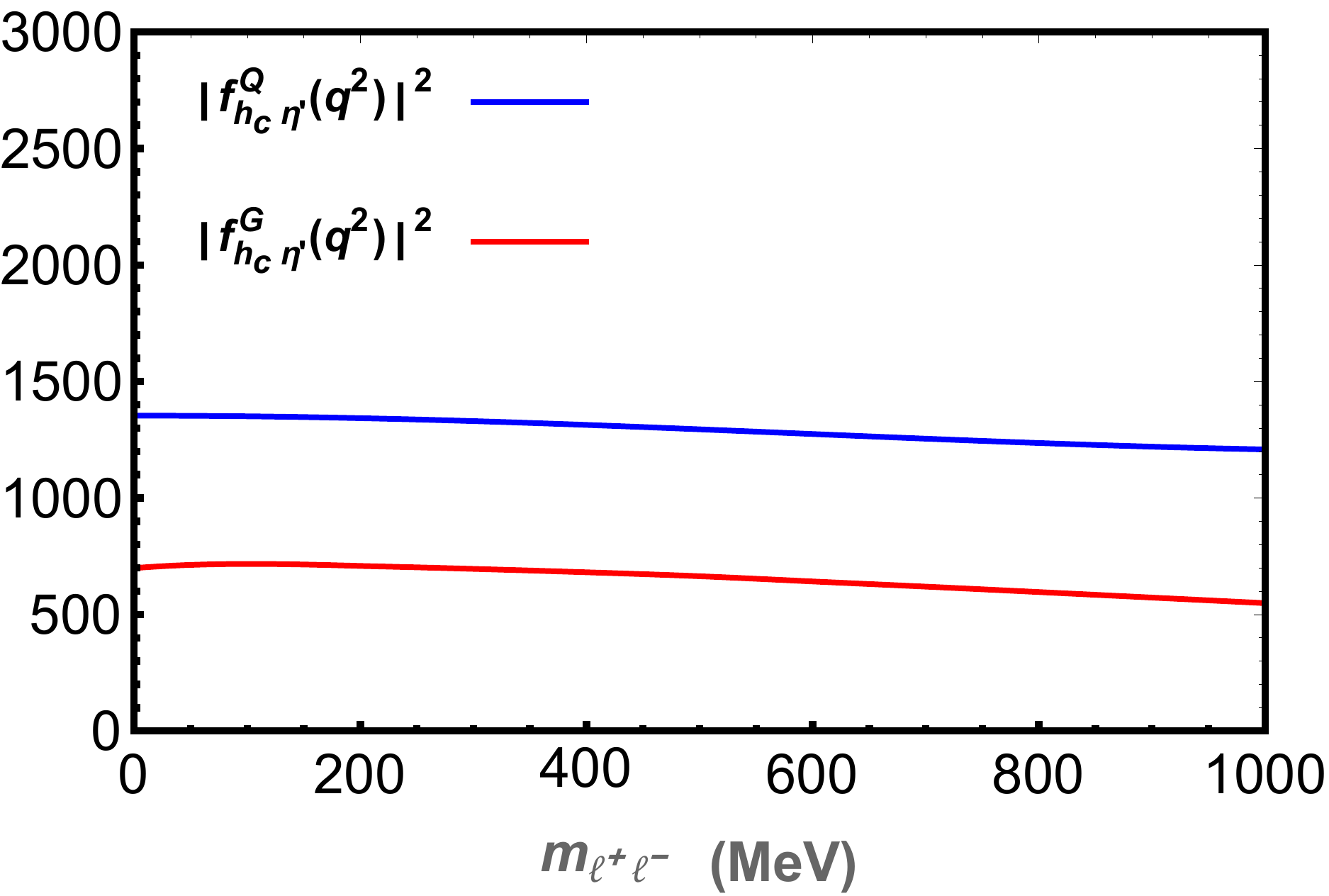}
\caption{\label{tffmll}The dependence of the modulus square of the TFFs $|f^{Q,G}_{h_{c} \eta^{(\prime)}}(q^{2})|^{2}$ on the dilepton invariant mass $m_{\ell^{+}\ell^{-}}$ (or, $q^{2}$).}
\end{figure}

Performing the integral calculations of the TFFs $f^{Q}_{h_{c} \eta^{(\prime)}}(q^{2})$ and $f^{G}_{h_{c} \eta^{(\prime)}}(q^{2})$, we find that the modulus square of these TFFs is very insensitive to the dilepton invariant mass $m_{\ell^{+}\ell^{-}}$ (or, $q^{2}$). In Fig.~\ref{tffmll}, the $m_{\ell^{+}\ell^{-}}$ dependence of the modulus square $|f^{Q,G}_{h_{c} \eta^{(\prime)}}(q^{2})|^{2}$ is shown. Schematically, we can clearly see that the modulus square $|f^{Q,G}_{h_{c} \eta^{(\prime)}}(q^{2})|^{2}$ has only negligible changes in the range of $(0-1000)\, \mathrm{MeV}$. Comparing the quark-antiquark contributions from $|f^{Q}_{h_{c} \eta^{(\prime)}}(q^{2})|^{2}$ with the gluonic contributions from $|f^{G}_{h_{c} \eta^{(\prime)}}(q^{2})|^{2}$, we find that the former is about twice that of the latter. In other words, the gluonic contributions and the quark-antiquark contributions are both important in these decay processes. It is compatible with the situation in the corresponding radiative decays $h_{c} \rightarrow\gamma\eta^{(\prime)}$~\cite{Fan:2019sap,He:2020kin}.

Based on the foregoing discussions, in the large recoil momentum region of $\eta^{(\prime)}$, the $h_{c}\rightarrow \eta^{(\prime)}\gamma^{\ast}$ TFFs can be obtained by
\begin{eqnarray}
f^{H}_{h_{c} \eta^{(\prime)}}(q^{2})=f^{Q}_{h_{c} \eta^{(\prime)}}(q^{2})+f^{G}_{h_{c} \eta^{(\prime)}}(q^{2}),
\end{eqnarray}
which includes the dynamical structure information from the quark-antiquark content and the gluonic content of $\eta^{(\prime)}$. In addition, the relativistic corrections related to the internal momentum of $h_{c}$ are taken into account in the TFFs. Specifically, there exist the kinematical corrections from the annihilation amplitudes and the dynamical corrections from the bound-state wave function of $h_{c}$. Since the physical picture in the large recoil momentum region of the electromagnetic Dalitz decays $h_{c}\rightarrow\eta^{(\prime)}\ell^{+}\ell^{-}$ is the same with the one in the corresponding radiative decays, these electromagnetic Dalitz decays can be calculated by the non-relativistic quark model with zero-binding approximation, as well as the Bethe-Salpeter formalism. Similarly to the situations in the radiative decays $h_{c}\rightarrow\eta^{(\prime)}\gamma$~\cite{Fan:2019sap,He:2020kin}, we find that these relativistic corrections also play an important role in the decay processes $h_{c}\rightarrow\eta^{(\prime)}\ell^{+}\ell^{-}$. Numerically, there is about a 2-times enhancement of $|f^{H}_{h_{c} \eta^{(\prime)}}(q^{2})|^{2}$ over the results with the zero-binding approximation.

\subsection{Soft mechanism}
\label{subsec:Soft mechanism}
Generally speaking, a perturbative QCD approach will become invalid in the three-body decay processes $h_{c}\rightarrow\eta^{(\prime)}\ell^{+}\ell^{-}$ with a small recoil momentum of $\eta^{(\prime)}$, and a special handling is needed in principle. To deal with this issue properly, a picture of the soft wave function overlap is proposed in Ref.~\cite{He:2020jvj}, where the picture has been proved valid in the decays $J/\psi\rightarrow\eta e^{+} e^{-}$ with a small recoil momentum. Because of the similar physical picture, the TFFs of the $h_{c}\rightarrow\eta^{(\prime)}\gamma^{\ast}$ in the small recoil momentum region, which are dependent on the recoil momentum $|\mathbf{p}_{\eta^{(\prime)}}|$ and reflect the size effects from the spatial wave functions of the initial- and final-state hadrons, can be adopted the empirical form in the rest frame of $h_{c}$~\cite{Amsler:1995td,Close:2000yk,Dudek:2006ej,Li:2007ky}:
\begin{eqnarray}
f^{S}_{h_{c} \eta^{(\prime)}}(q^{2})=g_{h_{c}  \eta^{(\prime)}}\exp\left(-\frac{\boldsymbol{q}^{2}}{8\beta^{2}}\right),
\end{eqnarray}
Here $\boldsymbol{q}^{2}=|\mathbf{p}_{\eta^{(\prime)}}|^{2}=\lambda(M^{2}, \, m^{2}, \, q^{2})/(4M^{2})$, $g_{h_{c}  \eta^{(\prime)}}$ denote the $h_{c} -\eta^{(\prime)}-\gamma^{\ast}$ coupling and can be determined by the continuity condition of the TFFs between the large and the small recoil momentum regions, and the parameter $\beta$ is an experiment-related quantity. We adopt $\beta = 400\, \mathrm{MeV}$, which is compatible with the fitted value in the $J/\psi$ decays~\cite{He:2020jvj}. And it is worth noting that the decay amplitude $\mathcal{A}^{\alpha\beta}$ in soft mechanism has the form $e f^{S}_{h_{c} \eta^{(\prime)}}(q^{2})\left(g^{\alpha\beta}-q^{\alpha}K^{\beta}/q\cdot K\right)$, in which the spin structure is determined just by the quantum number of the initial- and final-state hadrons.

In the whole recoil momentum region of $\eta^{(\prime)}$, the TFFs can be given by
\begin{eqnarray}
f_{h_{c} \eta^{(\prime)}}(q^{2})=
\begin{cases}
f^{H}_{h_{c} \eta^{(\prime)}}(q^{2})      ~~~~& q^{2}\leq 1\, \mathrm{GeV}^{2} , \\
f^{S}_{h_{c} \eta^{(\prime)}}(q^{2})      ~~~~& q^{2}>1\, \mathrm{GeV}^{2} .
\end{cases}
\end{eqnarray}
Incidentally, the recoil momentum $|\mathbf{p}_{\eta^{(\prime)}}|=\lambda^{\frac{1}{2}}(M^{2}, \, m^{2}, \, q^{2})/(2M)$ is a monotonically decreasing function of the square of the invariant mass of the lepton pair $q^{2}$. The recoil momentum is above $1\, \mathrm{GeV}$ when $q^{2}\leq 1\, \mathrm{GeV}^{2}$. It is commonly asserted that perturbative QCD is self-consistent when the recoil momentum is above $1\, \mathrm{GeV}$~\cite{Radyushkin:1990te,Jakob:1993iw,Jakob:1994hd,Bolz:1997ez}. Namely, the transition to perturbative QCD appears at about $q^{2}= 1\, \mathrm{GeV}^{2}$, and the hard mechanism begins to dominate as the $q^{2}$ decreases. On the contrary, the contributions from the soft mechanism would become important with the $q^{2}$ increases. Although we could obtain the hard contributions from the large recoil momentum region with the perturbative QCD approach and the soft ones from the small recoil momentum region with the overlapping integration of the soft wave functions, how to precisely match these two contributions in the intermediate recoil momentum region is still an open question and needs further investigation. Even so, our description of the EM Dalitz decay processes $h_{c}\rightarrow\eta^{(\prime)}\ell^{+}\ell^{-}$ may constitute an important step forward toward a satisfactory description.

\section{RESULTS AND DISCUSSIONS}
\label{sec:Results and discussions}

In the rest frame of $h_{c}$, the $q^{2}$-dependent differential decay widths of $h_{c}\rightarrow\eta^{(\prime)}\ell^{+}\ell^{-}$ can be written as
\begin{eqnarray}\label{unnormalizedwid}
\frac{\mathrm{d}\Gamma(h_{c}\rightarrow\eta^{(\prime)}\ell^{+}\ell^{-})}{\mathrm{d}q^{2}}&=&\frac{\alpha^{2}}{18\pi}\frac{\lambda^{\frac{1}{2}}(M^{2}, \, m^{2}, \, q^{2})}{M^{3}}\frac{\mid f_{h_{c} \eta^{(\prime)}}(q^{2}) \mid^{2}}{q^{2}}\left(1+\frac{2m_{\ell}^{2}}{q^{2}}\right)\nonumber\\
&&\times\left(1-\frac{4m_{\ell}^{2}}{q^{2}}\right)^{\frac{1}{2}}\left(1+\frac{2M^{2}q^{2}}{(M^{2}-m^{2}+q^{2})^{2}}\right),
\end{eqnarray}
where $m_{\ell}$ is the lepton mass. In order to remove most of the uncertainties from the TFFs (a brief discussion in what follows), we relate the differential decay widths $\mathrm{d}\Gamma(h_{c}\rightarrow\eta^{(\prime)}\ell^{+}\ell^{-})$ to the corresponding radiative decay widths $\Gamma(h_{c}\rightarrow\eta^{(\prime)}\gamma)$:
\begin{eqnarray}\label{normalizedwid}
\frac{\mathrm{d}\Gamma(h_{c}\rightarrow\eta^{(\prime)}\ell^{+}\ell^{-})}{\mathrm{d}q^{2}\Gamma(h_{c}\rightarrow\eta^{(\prime)}\gamma)}
&=&\frac{\alpha}{3\pi}\mid F_{h_{c} \eta^{(\prime)}}(q^{2}) \mid^{2}\frac{1}{q^{2}}\frac{\lambda^{\frac{1}{2}}(M^{2}, \, m^{2}, \, q^{2})}{(M^{2}-m^{2})}\left(1+\frac{2m_{\ell}^{2}}{q^{2}}\right)\nonumber\\
&&\times\left(1-\frac{4m_{\ell}^{2}}{q^{2}}\right)^{\frac{1}{2}}\left(1+\frac{2M^{2}q^{2}}{(M^{2}-m^{2}+q^{2})^{2}}\right),
\end{eqnarray}
where $F_{h_{c} \eta^{(\prime)}}(q^{2})\equiv f_{h_{c} \eta^{(\prime)}}(q^{2}) /f_{h_{c} \eta^{(\prime)}}(0)$ are the normalized TFFs, and the normalization is such that $F_{h_{c} \eta^{(\prime)}}(0)=1$.

In the numerical calculations, all the values of the involved meson masses, quark masses, decay widths and decay constant are quoted from the Particle Data Group~\cite{ParticleDataGroup:2022pth}. By employing the two-loop renormalization group equation, we obtain the strong coupling constant $\alpha_{s}(m_{c})=0.38$. The effective mass of the $c$ quark and the harmonic oscillator parameter appearing in the bound-state wave function are taken as $\hat{m}_{c}=1490~\mathrm{MeV}$ and $\beta_{A}=590~\mathrm{MeV}$ respectively, and more discussions can be found in Refs.~\cite{Negash:2015rua,Bhatnagar:2016otj,Gebrehana:2019mpw}. For the Gegenbauer moments from $\eta^{(\prime)}$ DAs, we just adopt model I in Table~1 of Ref.~\cite{Fan:2019sap} due to the negligibly small uncertainties from these Gegenbauer moments, which have been mentioned in the foregoing discussions. For the phenomenological parameters, i.e., the mixing angle $\phi$ and the decay constants $f_{q(s)}$, we adopt the set of values~\cite{Escribano:2013kba}
\begin{eqnarray}
\phi=33.5^{\circ}\pm 0.9^{\circ},~~~~f_{q}=(1.09\pm 0.02)f_{\pi},~~~~f_{s}=(0.96\pm 0.04)f_{\pi}
\end{eqnarray}
extracted from the TFF $F_{\gamma^{\ast}\gamma\eta^{\prime}}(+\infty)$, which is in excellent agreement with the \textsl{BABAR} measurement~\cite{Aubert:2006cy}. And more discussions about these phenomenological parameters can be found in Refs.~\cite{Gregory:2011sg,Chen:2014yta,He:2019mpy,Fan:2019sap,He:2020kin,Jiang:2022gnd}.

\begin{table}[!htbp]
  \caption{\label{tab:HS}The branching ratios $\mathcal{B}(h_{c}\rightarrow\eta^{(\prime)}\ell^{+}\ell^{-})$.}
  \vspace{0.2cm}
  \centering
  \begin{tabular}{lccc}
  \hline\hline
  ~~&~~Hard mechanism~~~ &~~Soft mechanism~~&~~Total~~~  \\
  \hline
 $\mathcal{B}(h_{c}\rightarrow\eta e^{+}e^{-})$~~&~~$4.9\times10^{-6}$~~&~~$3.7\times10^{-6}$~~&~~$8.5\times10^{-6}$ \\
 $\mathcal{B}(h_{c}\rightarrow\eta \mu^{+}\mu^{-})$~~&~~$1.0\times10^{-6}$~~&~~$3.7\times10^{-6}$~~&~~$4.6\times10^{-6}$ \\
 $\mathcal{B}(h_{c}\rightarrow\eta^{\prime}e^{+}e^{-})$~~&~~$1.6\times10^{-5}$
 ~~&~~$0.7\times10^{-5}$~~&~~$2.3\times10^{-5}$\\
 $\mathcal{B}(h_{c}\rightarrow\eta^{\prime}\mu^{+}\mu^{-})$~~&~~$0.3\times10^{-5}$
 ~~&~~$0.7\times10^{-5}$~~&~~$1.0\times10^{-5}$\\
  \hline\hline
  \end{tabular}
\end{table}

We now proceed with a full calculation of the branching ratios $\mathcal{B}(h_{c}\rightarrow\eta^{(\prime)}\ell^{+}\ell^{-})$, and our numerical results are shown in Table~\ref{tab:HS}. Here we do not present the theoretical uncertainties which come mainly from $\mathcal{B}^{exp}(h_{c}\rightarrow\eta\gamma)=(4.7\pm 2.1)\times10^{-4}$ or $\mathcal{B}^{exp}(h_{c}\rightarrow\eta^{\prime}\gamma)=(1.5\pm 0.4)\times10^{-3}$~\cite{Ablikim:2016uoc}, and they are expected to be $30\% \sim 50\%$. In the second column, the results from the hard mechanism, which can be described by the perturbative QCD approach, are presented; in the third column, we show the results from the soft mechanism, which is governed by the overlapping integration of the wave functions of the initial- and final-state hadrons. And the total contributions from both the hard mechanism and the soft mechanism are presented in the last column. First of all, it is noticed that the soft contributions in the decay processes $h_{c}\rightarrow\eta^{(\prime)} e^{+} e^{-}$ are equal to the ones in the decay processes $h_{c}\rightarrow\eta^{(\prime)} \mu^{+} \mu^{-}$ with an accuracy of more than three (two) significant digits. This is because the differential branching ratios are proportional to $(1-{\mathcal O}(m_{\ell}^{4}/q^{4}))$ in large $q^{2}$ region (see Eq.~\eqref{unnormalizedwid} or Eq.~\eqref{normalizedwid}). Therefore, the difference caused by the lepton mass $m_{\ell}$ is ignorable in the small recoil momentum region. While, the hard contributions of the $\mu^{+}\mu^{-}$ channel are about five times smaller than the ones of the $e^{+}e^{-}$ channel, since the phase space of the former shrinks in the small $q^{2}$ region. Besides, we find that the hard contributions and the soft ones are comparable with each other. It is unlike the situation where the soft contributions are negligibly small in the $S$-wave charmonium EM Dalitz decays $J/\psi\rightarrow\eta^{(\prime)}\ell^{+}\ell^{-}$ because of a suppression of the kinematic factor (i.e., $|\mathbf{p}_{\eta^{(\prime)}}|^{3}/q$)~\cite{He:2020jvj}. It is worthwhile to point out that there are around $3$ billion $\psi(2S)$ events collected with the BESIII detector so far~\cite{BESIII:2020nme,Ablikim:2023pjg,BESIII:2023lcc}, and this may imply that our predictions of the branching ratios $\mathcal{B}(h_{c}\rightarrow\eta^{(\prime)}\ell^{+}\ell^{-})$ may come within the range of measurement of present or near-future experiments, especially for the $\eta^{\prime}$ channels (reaching $10^{-5}$).

\begin{figure}[!!htb]
\centering
\includegraphics[width=0.43\textwidth]{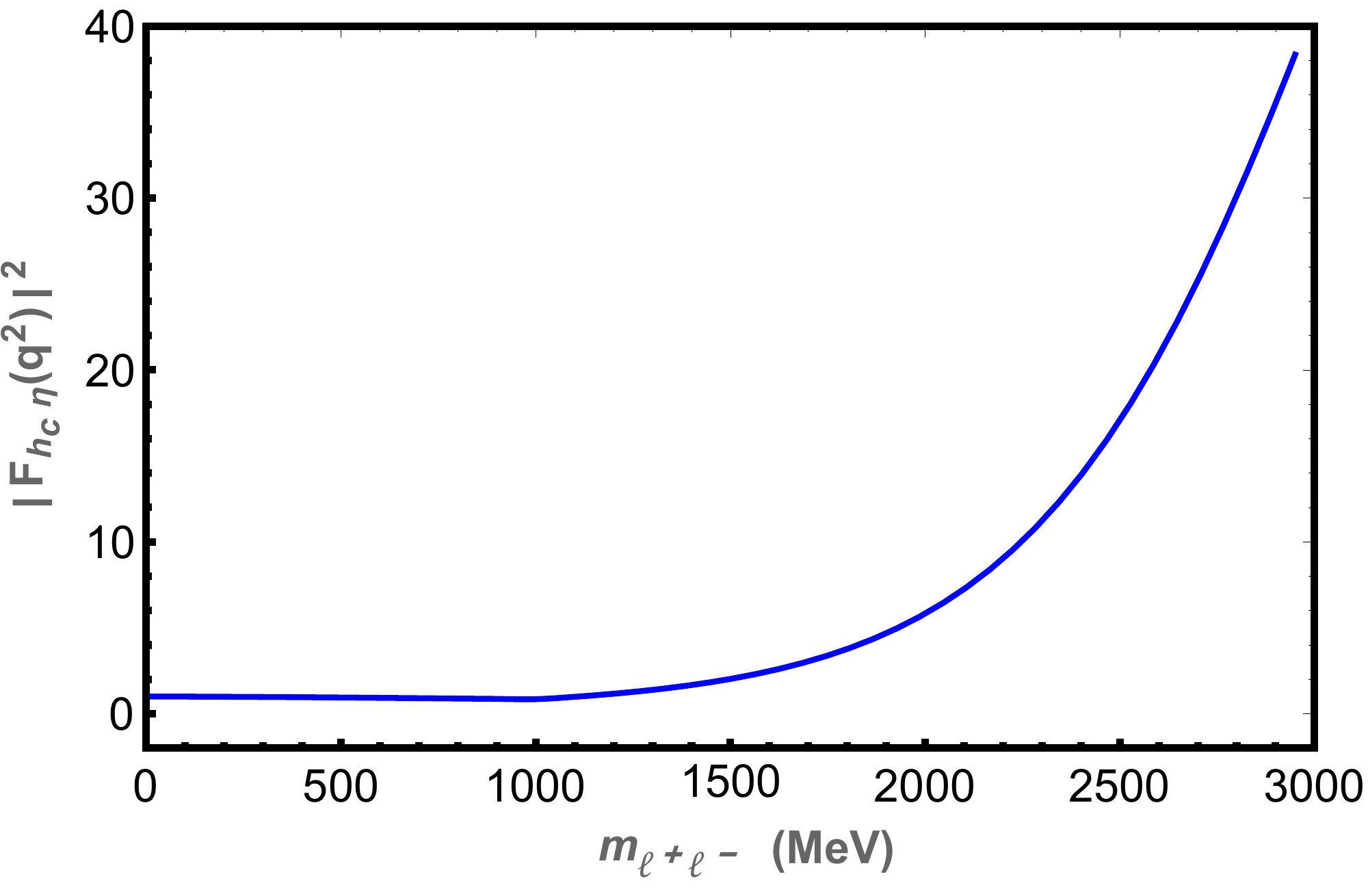}\hspace{0.5cm}
\includegraphics[width=0.42\textwidth]{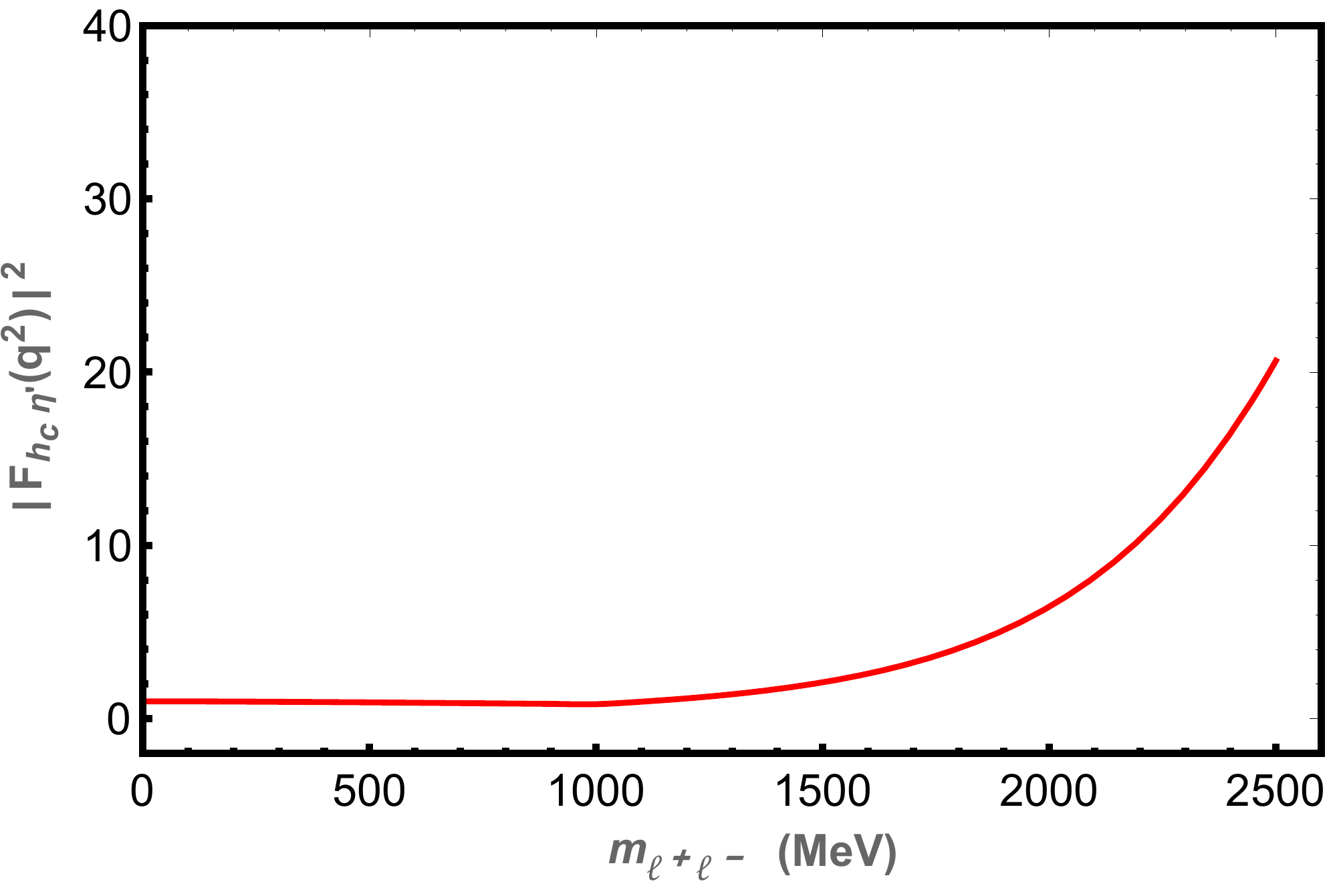}
\caption{\label{TFFHS}The dependence of the modulus square of the normalized TFFs $|F_{h_{c} \eta^{(\prime)}}(q^{2})|^{2}$ on the dilepton invariant mass $m_{\ell^{+}\ell^{-}}$ (or, $q^{2}$).}
\end{figure}

Last but not least, we concentrate on the $q^{2}$-dependent TFFs $f_{h_{c} \eta^{(\prime)}}(q^{2})$, which could provide the dynamical information of the EM structure arising at the $h_{c} \rightarrow \eta^{(\prime)}$ transition vertex and offer a powerful probe of the intrinsic structure of the $P$-wave charmonium $h_{c}$. Experimentally, one is interested in the normalized TFFs $F_{h_{c} \eta^{(\prime)}}(q^{2})$, because their modulus square $|F_{h_{c} \eta^{(\prime)}}(q^{2})|^{2}$ can be directly extracted by comparing the measured invariant mass spectrum of the lepton pairs from the Dalitz decays with the point-like QED prediction~\cite{Landsberg:1986fd,Fu:2011yy}. On the other hand, there exists a large uncertainty from sources such as the bound-state wave function and the QCD running coupling constant $\alpha_{s}$ in the TFFs $f_{h_{c} \eta^{(\prime)}}(q^{2})$. However, the dependence of the normalized TFFs $F_{h_{c} \eta^{(\prime)}}(q^{2})$ (i.e., the ratios $f_{h_{c} \eta^{(\prime)}}(q^{2}) /f_{h_{c} \eta^{(\prime)}}(0)$) on the bound-state wave function and the QCD running coupling constant is cut down to a large extent. So, one can expect that the predictions of the normalized TFFs are more reliable. Besides, the $q^{2}$ dependence of the TFFs $f_{h_{c} \eta^{(\prime)}}(q^{2})$ is still retained in the normalized TFFs $F_{h_{c} \eta^{(\prime)}}(q^{2})$ because of the constants $f_{h_{c} \eta^{(\prime)}}(0)$. In Fig.~\ref{TFFHS}, we present the $q^{2}$ dependence of the modulus square of the normalized TFFs $|F_{h_{c} \eta^{(\prime)}}(q^{2})|^{2}$ in their full kinematic region. Here it should be noted that, even though the normalized TFFs $|F_{h_{c} \eta^{(\prime)}}(q^{2})|^{2}$ are independent of the lepton mass, the kinematic region of the $e^{+}e^{-}$ channel is different from the one of the $\mu^{+}\mu^{-}$ channel. Specifically, the value of the dilepton invariant mass $m_{\ell^{+}\ell^{-}}$ starts at $2m_{e}$ in the $e^{+}e^{-}$ channel and it starts at $2m_{\mu}$ in the $\mu^{+}\mu^{-}$ channel. As shown schematically in Fig.~\ref{TFFHS}, the difference in the modulus squares $|F_{h_{c} \eta}(q^{2})|^{2}$ and $|F_{h_{c} \eta^{\prime}}(q^{2})|^{2}$ mainly arise from their phase space. One can find that the modulus square $|F_{h_{c} \eta^{(\prime)}}(q^{2})|^{2}$ is quite steady in small $q^{2}$ region and increasing rapidly in large $q^{2}$ region, which is compatible with the situation in the EM Dalitz decay processes $J\psi\rightarrow\eta^{(\prime)}\ell^{+}\ell^{-}$~\cite{BESIII:2014dax,He:2020jvj}. Present or near future experimental measurement is expected to provide tests for these predictions.

\section{SUMMARY}
\label{sec:conclusion}

In this paper, we investigate the $P$-wave charmonium EM Dalitz decays $h_{c}\rightarrow\eta^{(\prime)}\ell^{+}\ell^{-}$ with a QCD analysis. In the large recoil momentum region of $\eta^{(\prime)}$, these decay processes are described by the perturbative QCD approach. For the primary heavy meson $h_{c}$, we work out its B-S wave function in the framework of B-S equation, and its internal momentum is retained in both the wave function and the hard-scattering amplitude; for the final light mesons $\eta^{(\prime)}$, the light-cone DAs are adopted due to a large momentum transfer. By an analytic calculation of the involved one-loop integrals, we find that the TFFs are UV and IR safe, and the gluonic contributions and the quark-antiquark contributions are both important in the TFFs. In the small recoil momentum region of $\eta^{(\prime)}$, the picture of the soft wave function overlap is adopted to describe the transition mechanism of $h_{c} \to \eta^{(\prime)}$. By relating to their radiative decay processes, the branching ratios $\mathcal{B}(h_{c}\rightarrow\eta^{(\prime)}\ell^{+}\ell^{-})$ are obtained. Intriguingly, the contributions from the soft mechanism and those from the hard mechanism are comparable with each other, unlike the situation in $S$-wave charmonium decays $J/\psi\rightarrow\eta^{(\prime)}\ell^{+}\ell^{-}$~\cite{He:2020jvj} where the soft contributions are suppressed because of the special form of the spin structure of their amplitudes. Furthermore, the $q^{2}$-dependent TFFs are analyzed briefly, and we obtain the $q^{2}$ dependence of the modulus square of the normalized TFFs $|F_{h_{c} \eta^{(\prime)}}(q^{2})|^{2}$ in their full kinematic region. Lastly, it should be pointed out that there are around $3$ billion $\psi(2S)$ events collected with the BESIII detector so far~\cite{BESIII:2020nme,Ablikim:2023pjg,BESIII:2023lcc}, and this may imply that our predictions of the branching ratios $\mathcal{B}(h_{c}\rightarrow\eta^{(\prime)}\ell^{+}\ell^{-})$ may come within the range of measurement of present or near-future experiments, especially for the $\eta^{\prime}$ channels.

\section*{ACKNOWLEDGMENTS}
This work is supported by the Guiding Project of Science and Technology Research Program of the Hubei Provincial Department of Education (Grant No. B2022160).

\newpage


\appendix

\section*{APPENDIX: THE EXPRESSIONS OF THE NINE NUMERATORS}
\label{sec:nine numerators}
The expressions of the numerators $N_{i}$ ($i=1\sim9$) read
\begin{eqnarray*}
N_{1}&=&\frac{8i f(\hat{k}^{2}) }{(M^{4}+m^{4}+q^{4}-2m^{2}(M^{2}+q^{2})+4M^{2}q^{2})}\frac{(M^{2}-m^{2}+q^{2})}{(M^{2}-2m^{2}-2q^{2}+4m^{2}_{c}-8\hat{k}\cdot p-4\hat{k}^2)}\\
&&\times\frac{1}{\hat{m}_{c}M}\Bigg{[} 2 \bigg{(}2 k_{1}^2 (\hat{k}^2 (m_{c}+\hat{m}_{c}) ((m^2-q^2)^2-M^4)+2 M^2 \hat{k}\cdot p (2 \hat{k}\cdot p (m_{c}-\hat{m}_{c})-\hat{m}_{c}\\
&&\times (m^2+2 M m_{c}+q^2)))+k_{1}\cdot p (4 k_{1}\cdot q (\hat{k}^2 (m_{c}+\hat{m}_{c}) (M^2-m^2+q^2)+\hat{m}_{c} \hat{k}\cdot p\\
&&\times (2 M m_{c}-m^2+q^2))+\hat{m}_{c} (m^2-M^2-q^2) (\hat{k}\cdot p (4 \hat{k}^2+4 \hat{k}\cdot p +3 m^2-2 M^2-4 m_{c}^2\\
&&+q^2)+2 \hat{k}^2 (m^2-M^2+q^2)))+\hat{m}_{c} k_{1}\cdot q (m^2-M^2-q^2) (\hat{k}\cdot p (4 \hat{k}^2+4 \hat{k}\cdot p+3 m^2 \\
&&-4 m_{c}^2+q^2) +4 \hat{k}^2 m^2)+(k_{1}\cdot p)^2 (4 \hat{k}^2 (m_{c}+\hat{m}_{c})(M^2-m^2+q^2) +4 \hat{m}_{c} \hat{k}\cdot p ( q^2\\
&&+2 M (M+m_{c})-m^2))\bigg{)}-k_{1}\cdot \hat{k} \bigg{(}\hat{m}_{c} (m^2-M^2-q^2) (4 \hat{k}^2 (m^2+M^2-q^2)+4 \hat{k}\cdot p \\
&&\times(m^2+M^2-q^2)+3 m^4-m^2 (M^2+4 m_{c}^2+2 q^2)+(M^2-q^2) (q-2 m_{c}) (2 m_{c}+q))\\
&&-4 k_{1}\cdot p (4 M^2 \hat{k}\cdot p (\hat{m}_{c}-m_{c})+\hat{m}_{c} (m^2-M^2-q^2) (m^2-2 M m_{c}-q^2))\bigg{)}\Bigg{]},
\end{eqnarray*}
\begin{eqnarray*}
N_{2}&=&\frac{8i f(\hat{k}^{2}) }{(M^{4}+m^{4}+q^{4}-2m^{2}(M^{2}+q^{2})+4M^{2}q^{2})}\frac{(M^{2}-m^{2}+q^{2})}{(M^{2}-2m^{2}-2q^{2}+4m^{2}_{c}+8\hat{k}\cdot p-4\hat{k}^2)}\\
&&\times\frac{1}{\hat{m}_{c}M}\Bigg{[} 2 \bigg{(}2 k_1^2 (\hat{k}^2 (m_{c}+\hat{m}_{c}) (m^4-2 m^2 q^2-M^4+q^4)+2 M^2 \hat{m}_{c} \hat{k}\cdot p (m^2+2 M m_{c}\\
&&+q^2)+4 M^2 (\hat{k}\cdot p)^2 (m_{c}-\hat{m}_{c}))+\hat{m}_{c} k_1\cdot q (m^2-M^2-q^2) (-\hat{k}\cdot p (4 \hat{k}^2+3 m^2-4 m_{c}^2\\
&&+q^2)+4 \hat{k}^2 m^2+4 (\hat{k}\cdot p)^2)+k_1\cdot p (\hat{m}_{c} (m^2-M^2-q^2) (-\hat{k}\cdot p (4 \hat{k}^2+3 m^2-2 M^2\\
&&-4 m_{c}^2+q^2)+2 \hat{k}^2 (m^2-M^2+q^2)+4 (\hat{k}\cdot p)^2)-4 k_1\cdot q (\hat{k}^2 (m_{c}+\hat{m}_{c}) (m^2-M^2-q^2)\\
&& +\hat{m}_{c} \hat{k}\cdot p (2 M m_{c}-m^2+q^2)))-4 (k_1\cdot p)^2 (\hat{k}^2 (m_{c}+\hat{m}_{c}) (m^2-M^2-q^2)+\hat{m}_{c} \hat{k}\cdot p \\
&&\times(2 M^2-m^2+2 M m_{c}+q^2))\bigg{)} +k_1\cdot \hat{k} \bigg{(} \hat{m}_{c} (m^2-M^2-q^2) (4 \hat{k}^2 (m^2+M^2-q^2)\\
&&-4 \hat{k}\cdot p (m^2+M^2-q^2)+3 m^4-m^2 M^2-4 m^2 m_{c}^2-2 m^2 q^2-4 M^2 m_{c}^2+M^2 q^2+4 m_{c}^2 q^2\\
&& -q^4)-4 k_1\cdot p (4 M^2 \hat{k}\cdot p (m_{c}-\hat{m}_{c})+\hat{m}_{c} (m^2-M^2-q^2) (m^2-2 M m_{c}-q^2))\bigg{)} \Bigg{]},
\end{eqnarray*}
\begin{eqnarray*}
N_{3}&=&\frac{8i f(\hat{k}^{2}) }{(M^{4}+m^{4}+q^{4}-2m^{2}(M^{2}+q^{2})+4M^{2}q^{2})}\frac{(M^{2}-m^{2}+q^{2})}{(M^{2}-2m^{2}-2q^{2}+4m^{2}_{c}-8\hat{k}\cdot p-4\hat{k}^2)}\\
&&\times\frac{1}{\hat{m}_{c}M}\Bigg{[} 4 k_1^2 \hat{k}^2 (m_{c}+\hat{m}_{c}) ((m^2-q^2)^2-M^4)+2 \hat{m}_{c} \hat{k}\cdot p \bigg{(}k_1\cdot q (4 \hat{k}^2 (M^2-m^2+q^2)\\
&&+m^4+m^2 (3 M^2-8 M m_{c}+4 m_{c}^2-2 q^2)-(M^2+q^2) (4 m_{c}^2-q^2))-4 k_1^2 M^2 (m^2+2 M m_{c}\\
&&+q^2)\bigg{)}+8 (\hat{k}\cdot p)^2 \bigg{(}2 k_1^2 M^2 (m_{c}-\hat{m}_{c})+\hat{m}_{c} k_1\cdot q (M^2-m^2+q^2)\bigg{)}+2 k_1\cdot p \bigg{(}\hat{m}_{c} \hat{k}\cdot p \\
&&\times(4 k_1\cdot q (2 M m_{c}-m^2+q^2)+q^2 (4 \hat{k}^2-2 m^2+3 M^2-4 m_{c}^2)+(m^2-M^2) (-4 \hat{k}^2+m^2\\
&&+2 M^2-8 M m_{c}+4 m_{c}^2)+q^4)+2 \hat{k}^2 (m^2-M^2-q^2) (m_{c} (m^2-M^2+q^2)-2 k_1\cdot q (m_{c}\\
&&+\hat{m}_{c}))+4 (\hat{k}\cdot p)^2 (\hat{m}_{c} (q^2-m^2)+M^2 (3 \hat{m}_{c}-2 m_{c}))\bigg{)}+8 (k_1\cdot p)^2 \bigg{(}\hat{k}^2 (m_{c}+\hat{m}_{c}) (M^2\\
&&-m^2+q^2)+\hat{m}_{c} \hat{k}\cdot p (-m^2+2 M (M+m_{c})+q^2)\bigg{)}+8 \hat{k}^2 m^2 m_{c} k_1\cdot q (m^2-M^2-q^2)\\
&&+k_1\cdot \hat{k} \bigg{(}4 \hat{k}\cdot p (4 M^2 k_1\cdot p (\hat{m}_{c}-m_{c})+m^4 \hat{m}_{c}+m^2 (4 M^2 (m_{c}-\hat{m}_{c})-2 \hat{m}_{c} q^2)+\hat{m}_{c} \\
&&\times(q^4-M^4))-\hat{m}_{c} (m^2-M^2-q^2) (4 k_1\cdot p (2 M m_{c}-m^2+q^2)-4 \hat{k}^2 (m^2+M^2-q^2)\\
&&+m^4+m^2 (M^2-8 M m_{c}+4 m_{c}^2-2 q^2)-(M^2-q^2) (q^2-2 m_{c}^2))\bigg{)}\Bigg{]},
\end{eqnarray*}
\begin{eqnarray*}
N_{4}&=&\frac{8i f(\hat{k}^{2}) }{(M^{4}+m^{4}+q^{4}-2m^{2}(M^{2}+q^{2})+4M^{2}q^{2})}\frac{(M^{2}-m^{2}+q^{2})}{(M^{2}-2m^{2}-2q^{2}+4m^{2}_{c}+8\hat{k}\cdot p-4\hat{k}^2)}\\
&&\times\frac{1}{\hat{m}_{c}M}\Bigg{[} 4 k_1^2 \hat{k}^2 (m_{c}+\hat{m}_{c}) ((m^2-q^2)^2-M^4)+2 \hat{m}_{c} \hat{k}\cdot p \bigg{(} 4 k_1^2 M^2 (m^2+2 M m_{c}+q^2)\\
&&-k_1\cdot q (4 \hat{k}^2 (M^2-m^2+q^2)+m^4+m^2 (3 M^2-8 M m_{c}+4 m_{c}^2-2 q^2)-(M^2+q^2) \\
&&\times(4 m_{c}^2-q^2))\bigg{)}+8 (\hat{k}\cdot p)^2 \bigg{(} 2 k_1^2 M^2 (m_{c}-\hat{m}_{c})+\hat{m}_{c} k_1\cdot q (M^2-m^2+q^2)\bigg{)}+2 k_1\cdot p \\
&&\times\bigg{(} \hat{m}_{c} \hat{k}\cdot p (4 k_1\cdot q (m^2-2 M m_{c}-q^2)+q^2 (-4 \hat{k}^2+2 m^2-3 M^2+4 m_{c}^2)-(m^2-M^2)\\
&&\times (-4 \hat{k}^2+m^2+2 M^2-8 M m_{c}+4 m_{c}^2)-q^4)+2 \hat{k}^2 (m^2-M^2-q^2) (m_{c} (m^2-M^2+q^2)\\
&&-2 k_1\cdot q (m_{c}+\hat{m}_{c}))+4 (\hat{k}\cdot p)^2 (\hat{m}_{c} (q^2-m^2)+M^2 (3 \hat{m}_{c}-2 m_{c}))\bigg{)}+8 (k_1\cdot p)^2 \bigg{(} \hat{k}^2 \\
&&\times(m_{c}+\hat{m}_{c}) (M^2-m^2+q^2)+\hat{m}_{c} \hat{k}\cdot p (m^2-2 M (M+m_{c})-q^2)\bigg{)}+8 \hat{k}^2 m^2 m_{c} k_1\cdot q \\
&&\times(m^2-M^2-q^2)+k_1\cdot \hat{k} \bigg{(} \hat{m}_{c} (m^2-M^2-q^2) (4 k_1\cdot p (2 M m_{c}-m^2+q^2)-4 \hat{k}^2 (m^2\\
&&+M^2-q^2)+m^4+m^2 (M^2-8 M m_{c}+4 m_{c}^2-2 q^2)-(M^2-q^2) (q^2-4 m_{c}^{2}) )+4 \hat{k}\cdot p\\
&&\times (4 M^2 k_1\cdot p (\hat{m}_{c}-m_{c})+m^4 \hat{m}_{c}+m^2 (4 M^2 (m_{c}-\hat{m}_{c})-2 \hat{m}_{c} q^2)+\hat{m}_{c} (q^4-M^4))\bigg{)}\Bigg{]},
\end{eqnarray*}
\begin{eqnarray*}
N_{5}&=&\frac{2i f(\hat{k}^{2}) }{(M^{4}+m^{4}+q^{4}-2m^{2}(M^{2}+q^{2})+4M^{2}q^{2})}\frac{(M^{2}-m^{2}+q^{2})}{(M^{2}-2m^{2}-2q^{2}+4m^{2}_{c}-8\hat{k}\cdot p-4\hat{k}^2)}\\
&&\times\frac{1}{\hat{m}_{c}M}\Bigg{[} -8 k_1^2 \hat{k}^2 \hat{m}_{c} ((m^2-q^2)^2-M^4) +k_1\cdot \hat{k} \bigg{(} (m^2-M^2-q^2) (\hat{m}_{c} (-4 k_1^2 (m^2+M^2\\
&&-q^2)+m^4+m^2 (M^2-4 M m_{c}-4 m_{c}^2-2 q^2)+(M^2-q^2) (4 m_{c} (M-m_{c})-q^2))\\
&&+4 \hat{k}^2 (2 m_{c}-\hat{m}_{c}) (m^2+M^2-q^2))-4 \hat{k}\cdot p (4 (k_1\cdot p+k_1\cdot q) (m^2 (2 m_{c}-\hat{m}_{c})-M^2 \hat{m}_{c}\\
&&+q^2 (\hat{m}_{c}-2 m_{c}))+(m^2+M^2-q^2) (\hat{m}_{c} (m^2-q^2)+M^2 (\hat{m}_{c}-2 m_{c})))-8 \hat{m}_{c}\\
&&\times (m^2-2 k_1\cdot p) (k_1\cdot p+k_1\cdot q) (m^2-M^2-q^2)\bigg{)}-2 \hat{k}\cdot p \bigg{(} k_1\cdot p (\hat{m}_{c} (-4 k_1^2 (m^2+3 M^2\\
&&-q^2)+q^2 (-2 m^2+3 M^2-4 M m_{c}+4 m_{c}^2)+(m-M) (m+M) (m^2+2 M^2-4 M m_{c}\\
&&-4 m_{c}^2)+q^4)+16 \hat{m}_{c} k_1\cdot q (k_1\cdot q-m^2)+4 \hat{k}^2 (2 m_{c}-\hat{m}_{c}) (m^2-M^2-q^2))+k_1\cdot q \\
&&\times(\hat{m}_{c} (-4 k_1^2 (m^2+3 M^2-q^2)+m^4+m^2 ((M-2 m_{c}) (3 M+2 m_{c})-2 q^2)-(M^2+q^2)\\
&&\times (4 m_{c} (M-m_{c})-q^2))-8 m^2 \hat{m}_{c} k_1\cdot q+4 \hat{k}^2 (2 m_{c}-\hat{m}_{c}) (m^2-M^2-q^2))+4 k_1^2 M^2\\
&&\times \hat{m}_{c} (m^2+M^2-q^2)-8 \hat{m}_{c} (k_1\cdot p)^2 (m^2-4 k_1\cdot q)+16 \hat{m}_{c} (k_1\cdot p)^3\bigg{)}-4 \hat{k}^2 \hat{m}_{c} (m^2-M^2\\
&&-q^2) \bigg{(} k_1\cdot p (-4 k_1\cdot p+m^2-M^2+q^2)+2 k_1\cdot q (m^2-2 k_1\cdot p)\bigg{)}+8 m_{c} (k_1\cdot \hat{k})^2 ((m^2-q^2)^2\\
&&-M^4)+8 (\hat{k}\cdot p)^2 (k_1\cdot p+k_1\cdot q) \bigg{(} 4 (m_{c}-\hat{m}_{c}) (k_1\cdot p+k_1\cdot q)+\hat{m}_{c} (m^2-q^2)\\
&&+M^2 (\hat{m}_{c}-2 m_{c})\bigg{)}\Bigg{]},
\end{eqnarray*}
\begin{eqnarray*}
N_{6}&=&\frac{2i(M^{2}-m^{2}+q^{2}) f(\hat{k}^{2}) }{\hat{m}_{c}M(M^{4}+m^{4}+q^{4}-2m^{2}(M^{2}+q^{2})+4M^{2}q^{2})}\Bigg{[}-8 k_1^2 \hat{k}^2 \hat{m}_{c} ((m^2-q^2)^2-M^4)\\
&&+k_1\cdot \hat{k} \bigg{(}-(m^2-M^2-q^2) ( \hat{m}_{c} (-4 k_1^2 (m^2+M^2-q^2)+m^4+m^2 (M^2-4 M m_{c}-4 m_{c}^2\\
&&-2 q^2)+(M^2-q^2) (4 m_{c} (M-m_{c})-q^2))+4 \hat{k}^2 (2 m_{c}-\hat{m}_{c}) (m^2+M^2-q^2))-4 \hat{k}\cdot p\\
&&\times ( 4 (k_1\cdot p+k_1\cdot q) (m^2 (2 m_{c}-\hat{m}_{c})-M^2 \hat{m}_{c}+q^2 (\hat{m}_{c}-2 m_{c}))+(m^2+M^2-q^2)\\
&&\times (\hat{m}_{c} (m^2-q^2)+M^2 (\hat{m}_{c}-2 m_{c})))+8 \hat{m}_{c} (m^2-2 k_1\cdot p) (k_1\cdot p+k_1\cdot q) (m^2-M^2\\
&&-q^2)\bigg{)}+2 \hat{k}\cdot p \bigg{(} k_1\cdot p (\hat{m}_{c} (-4 k_1^2 (m^2+3 M^2-q^2)+q^2 (3 M^2-2 m^2-4 M m_{c}+4 m_{c}^2)\\
&&+(m^2-M^2) (m^2+2 M^2-4 M m_{c}-4 m_{c}^2)+q^4)+16 \hat{m}_{c} k_1\cdot q (k_1\cdot q-m^2)+4 \hat{k}^2 (2 m_{c}\\
&&-\hat{m}_{c}) (m^2-M^2-q^2))+k_1\cdot q (\hat{m}_{c} (-4 k_1^2 (m^2+3 M^2-q^2)+m^4+m^2 ((M-2 m_{c})\\
&&\times (3 M+2 m_{c})-2 q^2)-(M^2+q^2) (4 m_{c} (M-m_{c})-q^2))-8 m^2 \hat{m}_{c} k_1\cdot q+4 \hat{k}^2 (2 m_{c}\\
&&-\hat{m}_{c}) (m^2-M^2-q^2))+4 k_1^2 M^2 \hat{m}_{c} (m^2+M^2-q^2)-8 \hat{m}_{c} (k_1\cdot p)^2 (m^2-4 k_1\cdot q)\\
&&+16 \hat{m}_{c} (k_1\cdot p)^3\bigg{)}-4 \hat{k}^2 \hat{m}_{c} (m^2-M^2-q^2) \bigg{(} k_1\cdot p (-4 k_1\cdot p+m^2-M^2+q^2)+2 k_1\cdot q \\
&&\times(m^2-2 k_1\cdot p)\bigg{)}+8 m_{c} (k_1\cdot \hat{k})^2 ((m^2-q^2)^2-M^4)+8 (\hat{k}\cdot p)^2 (k_1\cdot p+k_1\cdot q) \bigg{(} 4 (m_{c}\\
&&-\hat{m}_{c}) (k_1\cdot p+k_1\cdot q)+\hat{m}_{c} (m^2-q^2)+M^2 (\hat{m}_{c}-2 m_{c})\bigg{)}\Bigg{]},
\end{eqnarray*}
\begin{eqnarray*}
N_{7}&=&\frac{i (M^{2}-m^{2}+q^{2}) f(\hat{k}^{2}) }{2 \hat{m}_{c}M(M^{4}+m^{4}+q^{4}-2m^{2}(M^{2}+q^{2})+4M^{2}q^{2})}\frac{1}{(M^{2}+m^{2}-q^{2}+\lambda^{\frac{1}{2}}(M^2,q^2,m^2))}\\
&&\times\frac{1}{\lambda^{\frac{1}{2}}(M^2,q^2,m^2)}\Bigg{[} \lambda(M^2,q^2,m^2) \bigg{(} M^2 ( \hat{k} \cdot p ) (4 u (\hat{m}_{c}-m_{c}) ( \hat{k} \cdot p )+\hat{m}_{c} (m^2 (2 u-1)\\
&&+M^2-4 M m_{c} u-2 q^2 u+q^2))- \hat{k}^{2}  (m^4-2 m^2 q^2-M^4+q^4) (m_{c} u+\hat{m}_{c} (u-1))\bigg{)} \\
&&+\lambda^{\frac{1}{2}}(M^2,q^2,m^2) \bigg{(} M^2 ( \hat{k} \cdot p ) (4 ( \hat{k} \cdot p ) (m^2 (\hat{m}_{c} (u-1)-m_{c} u)+M^2 (-m_{c} u+\hat{m}_{c} u+\hat{m}_{c})\\
&&+q^2 (m_{c} u+\hat{m}_{c} (-u)+\hat{m}_{c}))+\hat{m}_{c} (-2 m^4 (u-1)+m^2 (M^2 (2 u-3)+4 M m_{c} u-4 m_{c}^2)\\
&&+M^4-4 M^3 m_{c} u+M^2 (4 m_{c}^2-q^2 (2 u+1))+4 M m_{c} q^2 u+4 m_{c}^2 q^2+2 q^4 u-2 q^4))- \hat{k}^{2} \\
&&\times (m^2-M^2-q^2) ((m^4-2 m^2 (M^2+q^2)+(M^2-q^2)^2) (m_{c} u+\hat{m}_{c} (u-1))-4 M^2 \hat{m}_{c}\\
&&\times ( \hat{k} \cdot p ))\bigg{)}+M^2 ( \hat{k} \cdot p ) \bigg{(} 4 \hat{m}_{c}  \hat{k}^{2}  (m^4-2 m^2 q^2-M^4+q^4)-4 ( \hat{k} \cdot p ) (m^4 \hat{m}_{c}-2 m^2 (\hat{m}_{c}\\
&&\times (2 M^2 u+q^2)-2 M^2 m_{c} u)+\hat{m}_{c} (q^4-M^4))-\hat{m}_{c} (m^2-M^2-q^2) (m^4 (4 u-3)\\
&&+m^2 (M^2-8 M m_{c} u+4 m_{c}^2+2 q^2 (1-2 u))-(M^2-q^2) (q^2-4 m_{c}^2))\bigg{)}\Bigg{]},
\end{eqnarray*}
\begin{eqnarray*}
N_{8}&=&\frac{i (M^{2}-m^{2}+q^{2}) f(\hat{k}^{2}) }{2 \hat{m}_{c}M(M^{4}+m^{4}+q^{4}-2m^{2}(M^{2}+q^{2})+4M^{2}q^{2})}\frac{1}{(M^{2}+m^{2}-q^{2}+\lambda^{\frac{1}{2}}(M^2,q^2,m^2))}\\
&&\times\frac{1}{\lambda^{\frac{1}{2}}(M^2,q^2,m^2)}\Bigg{[} \lambda^{\frac{1}{2}}(M^2,q^2,m^2) \bigg{(} M^2 ( \hat{k} \cdot p ) (4 (2 u-1) ( \hat{k} \cdot p ) (\hat{m}_{c} (q^2-m^2)+M^2 (2 m_{c}\\
&&-\hat{m}_{c}))+\hat{m}_{c} (-m^2 (M^2 (1-2 u)^2-4 m_{c}^2+4 q^2 (u-1) u)+M^4 (1-2 u)^2-M^2 (4 m_{c}^2+q^2 \\
&&\times(8 u^2-8 u+3))+8 M m_{c} q^2-4 m_{c}^2 q^2+4 q^4 (u-1) u))- \hat{k}^{2}  (m^2-M^2-q^2) (4 M^2 (2 m_{c}\\
&&-\hat{m}_{c}) ( \hat{k} \cdot p )+\hat{m}_{c} (2 u-1) (m^4-2 m^2 (M^2+q^2)+(M^2-q^2)^2))\bigg{)}+\hat{m}_{c} \lambda(M^2,q^2,m^2)\\
&&\times \bigg{(} M^2 ( \hat{k} \cdot p ) (m^2 (1-2 u)^2+M^2 (1-2 u)^2-4 M m_{c}-q^2 (1-2 u)^2)-(2 u-1)  \hat{k}^{2}  (m^4\\
&&-2 m^2 q^2-M^4+q^4))\bigg{)}+M^2 ( \hat{k} \cdot p ) \bigg{(}-4 (2 u-1) (m^2+M^2-q^2) ( \hat{k} \cdot p ) (\hat{m}_{c} (m^2-q^2)\\
&&+M^2 (\hat{m}_{c}-2 m_{c}))-4  \hat{k}^{2}  (2 m_{c}-\hat{m}_{c}) (m^4-2 m^2 q^2-M^4+q^4)-\hat{m}_{c} (m^2-M^2-q^2)\\
&&\times (m^4 (1-2 u)^2+m^2 (M^2 (1-2 u)^2-4 M m_{c}-4 m_{c}^2-2 q^2 (2 u^2-2 u+1))\\
&&+(M^2-q^2) (4 M m_{c}-4 m_{c}^2-q^2))\bigg{)}\Bigg{]},
\end{eqnarray*}
\begin{eqnarray*}
N_{9}&=&\frac{i (M^{2}-m^{2}+q^{2}) f(\hat{k}^{2}) }{2 \hat{m}_{c}M(M^{4}+m^{4}+q^{4}-2m^{2}(M^{2}+q^{2})+4M^{2}q^{2})}\frac{1}{(M^{2}+m^{2}-q^{2}+\lambda^{\frac{1}{2}}(M^2,q^2,m^2))}\\
&&\times\frac{1}{\lambda^{\frac{1}{2}}(M^2,q^2,m^2)}\Bigg{[} \lambda(M^2,q^2,m^2) \bigg{(} M^2 ( \hat{k} \cdot p ) (\hat{m}_{c} (m^2 (1-2 u)+M^2+4 M m_{c} (u-1)\\
&&+q^2 (2 u-1))-4 (u-1) (m_{c}-\hat{m}_{c}) ( \hat{k} \cdot p ))- \hat{k}^{2}  (m^4-2 m^2 q^2-M^4+q^4) (m_{c} (u-1)\\
&&+\hat{m}_{c} u)\bigg{)} -\lambda^{\frac{1}{2}}(M^2,q^2,m^2) \bigg{(} \hat{k}^{2}  (m^2-M^2-q^2) ((m^4-2 m^2 (M^2+q^2)+(M^2-q^2)^2)\\
&&\times (m_{c} (u-1)+\hat{m}_{c} u)-4 M^2 \hat{m}_{c} ( \hat{k} \cdot p ))+M^2 ( \hat{k} \cdot p ) (4 ( \hat{k} \cdot p ) (m^2 (m_{c} (u-1)-\hat{m}_{c} u)\\
&&+M^2 (m_{c} (u-1)-\hat{m}_{c} (u-2))+q^2 (m_{c} (-u)+m_{c}+\hat{m}_{c} u))+\hat{m}_{c} (-2 m^4 u+m^2 (M^2 \\
&&\times(2 u+1)+4 M m_{c} (u-1)+4 m_{c}^2)-M^4-4 M^3 m_{c} (u-1)+M^2 (q^2 (3-2 u)-4 m_{c}^2)\\
&&+4 M m_{c} q^2 (u-1)-4 m_{c}^2 q^2+2 q^4 u))\bigg{)}+M^2 ( \hat{k} \cdot p ) \bigg{(} 4 \hat{m}_{c}  \hat{k}^{2}  (m^4-2 m^2 q^2-M^4+q^4)\\
&&+4 ( \hat{k} \cdot p ) (m^4 \hat{m}_{c}-2 m^2 (2 M^2 (u-1) (m_{c}-\hat{m}_{c})+\hat{m}_{c} q^2)+\hat{m}_{c} (q^4-M^4))+\hat{m}_{c} (m^2\\
&&-M^2-q^2) (m^4 (4 u-1)-m^2 (M^2+8 M m_{c} (u-1)+4 m_{c}^2+2 q^2 (2 u-1))\\
&&+(M^2-q^2) (q^2-4 m_{c}^2))\bigg{)}\Bigg{]}
\end{eqnarray*}
with $\lambda(a,\,b,\,c)\equiv a^{2}+b^{2}+c^{2}-2(ab+bc+ac)$ as the usual K\"{a}ll\'{e}n function.

\newpage

\bibliographystyle{JHEP}
\bibliography{hejk}

\end{document}